# High-Ti induced planar-fault transformation toward superlattice extrinsic stacking faults and microtwins in crept CoNi-based superalloys

Zhida Liang[1,*], Xiang Xu[3], Fengxian Liu[4], Xi Zhang[3,*], Xin Liu[2], Mingyang Zhang[2], Jing Zhang[5,*], Li Wang[6], Florian Pyczak[1], Yinan Cui[2,*]

1. Institute of Materials Physics, Helmholtz-Zentrum Hereon, Max-Planck-Strasse 1, Geesthacht 21502, Germany
2. Applied Mechanics Lab., School of Aerospace Engineering, Tsinghua University, Beijing 100084, China
3. Institute for Materials Science, University of Stuttgart, Pfaffenwaldring 55, 70569 Stuttgart, Germany
4. Department of Mechanics of Solids, Surfaces and Systems, Faculty of Engineering Technology, University of Twente, Drienerlolaan 5, 7522NB Enschede, Netherlands
5. Key Laboratory of MEMS of Ministry of Education, School of Integrated Circuits, Southeast University, Nanjing, China
6. State Key Laboratory of Powder Metallurgy, Central South university, 410083 Changsha, China

*Corresponding author: xi.zhang@imw.uni-stuttgart.de (Xi Zhang), zhida.liang@outlook.com (Zhida Liang), jizh@seu.edu.cn (Jing Zhang), cyn@mail.tsinghua.edu.cn (Yinan Cui)

**Abstract**

Controlling planar fault shearing mechanisms is key for improving the high-temperature creep performance of γ′-strengthened high-temperature superalloys. This work examines how the Ti concentration in $L1_2$-strengthened CoNi-based alloys affects planar fault formation during creep. Interrupted compressive creep tests were conducted at 1223 K under air with a constant load stress of 241 MPa. We found, for the first time, that high Ti additions shift the dominant γ′ shearing mode from antiphase boundaries (APBs) in Ti-free and low-Ti alloys to superlattice extrinsic stacking faults (SESFs). Systematic *ab initio* calculations show that in high-Ti alloys, the elevated APB energy renders APB-shearing mode unfavorable. Nevertheless, the SESF energy decreases relative to that in low-Ti compositions, and an increased ratio of complex intrinsic stacking fault (CISF) to SESF energy promote the transformation of high-energy CISFs into lower-energy SESFs. Chemical analysis using scanning transmission electron microscopy combined with energy-dispersive X-ray spectroscopy further reveals that, SESFs in high-Ti alloys are enriched in Ti, Mo and W, yet no grid-like ordering is observed. Together with the *ab initio* calculations, Mo and W additions in high Ti alloys could facilitate the transformation from $L1_2$ structure to low-energy $D0_{24}$ structure, indicating Mo and W segregation along SESFs is energetically favourable. Furthermore, the successive SESF thickening facilitates microtwinning in the absence of $D0_{24}$ ordering along SESFs, as an additional big carrier for creep strain. These new findings clarify the role of Ti in controlling planar fault shearing mechanisms, providing new insights for optimizing the creep performance of next-generation CoNi-based superalloys.

**Keywords:** Titanium, planar faults, superalloys, TEM

## 1. Introduction

The $\gamma'$-phase ($L1_2$ structure) strengthened superalloys with superior high-temperature strength are key materials for the hot sections of gas turbines used in power and propulsion systems. Their excellent high-temperature creep resistance mainly arises from the high-volume fraction of coherent $\gamma'$ precipitates, which effectively impede dislocation glide and climb during deformation (*Reed 2008, Chatterjee et al. 2021, Przybyla et al. 2010, Li et al. 2020*). The $\gamma'$-$Ni_3(Al,Ti)$ strengthened Ni-based superalloys, such as Udimet 720Li, PE16, and TWM, have been widely used in commercial applications that demand high proof stress at elevated temperatures (*Gu et al. 2006, Gu et al. 2009, Jones et al. 2014*). This strengthening effect is primarily attributed to the higher antiphase boundary (APB) energy and volume fraction of $Ni_3(Al,Ti)$ in Ni-based superalloys (*Raynor and Silcock 1970*). Previous studies (*Gu et al. 2006, Chen et al. 2022*) have shown that low-to-moderate Ti additions increase the $\gamma'$ volume fraction and the $\gamma'/\gamma$ lattice misfit, thereby enhancing creep resistance of Ni-based superalloys through inhibiting dislocation glide and climb.

Similarly, low-Ti additions in Co-based and CoNi-based superalloys provide comparable benefits, improving creep resistance not only through increased $\gamma'$ volume fraction (*Bauer et al. 2012, Neumeier et al. 2015, Xue et al. 2014*) but also by elevating the APB and superlattice intrinsic stacking fault (SISF) energies of $Co_3(Al,W)$ (*Mottura et al. 2012, Titus et al. 2016, Rhein et al. 2018*). However, the effect of high-Ti concentration on the creep resistance and the associated deformation mechanism of CoNi-based superalloys remains poorly understood. In particular, the effect of Ti/Al ratio on the activation of distinct planar faults shearing mechanisms during creep in CoNi-based superalloys has not been systematically investigated.

At high temperatures, the dominant shearing mechanism within the $\gamma'$ phase depends on whether the high-energy APBs or low-energy SFs are activated. When high-energy APBs are activated, channel dislocations choose to climb over the $\gamma'$ precipitates by pipe diffusion (*Zehl et al. 2025*), inducing knitting process to form dislocation networks (*Yashiro et al. 2006, Parsa et al. 2024*) and leading to improved creep resistance at high temperature (> 850 °C). Conversely, when low-energy SFs are activated, channel dislocations tend to shear the $\gamma'$ precipitates rather than climb over them, typically resulting in inferior creep performance (*Unocic et al. 2011*). Especially, the microtwins formed via superlattice extrinsic stacking faults (SESFs) thickening contribute much bigger creep strain than other planar faults, such as SISFs and APBs, accounting for the majority (73% - 96% contribution) of the total plastic strain in the primary and secondary creep stages (*Lenz et al. 2019, Barba et al. 2017*).

The SESF formation mechanism in Ni-based superalloys was firstly proposed by Kear et al. (*Kear et al. 1968*), who suggested that the combined glide of three Shockley partials of alternating sign could produce a $D0_{24}$-type structure corresponding to a two-layer SESF in an $L1_2$ precipitate. More recent studies (*Kolbe 2001, Smith et al. 2016, Kovarik et al. 2009, Smith et al. 2015, Karpstein et al. 2023*) have demonstrated that the SESF formation is from the stacking-fault reactions between two CISFs. Through subsequent elemental diffusion and atomic reordering (atoms reshuffling), the two-layer CISF transforms into a two-layer SESF. In addition, the atomic reordering and chemical segregation at SESFs can induce local phase transformations (LPTs), referred to $D0_{24}$ ordering. For instance, in high Ti+Nb+Ta content alloys (*Smith et al. 2016, Lilensten et al. 2021, Egan et al. 2022*), such as ME501 and RRHT5, enrichment of Co, Nb, Ta and W at SESFs stabilizes the local η phase ($Ni_3$(Ti,Nb,Ta), $D0_{24}$), which suppress the thickening of $D0_{24}$-type SESFs to become microtwins, thereby enhancing creep resistance. These chemical-segregation driven LPT processes have been characterized at the atomic scale and are known to play a key role in the creep behavior of several Ni-based alloy systems (*Smith et al. 2015, Karpstein et al. 2023, Lilensten et al. 2021*). In the low Ti+Nb+Ta content alloy ME03 and RRHT3 (*Smith et al. 2016, Titus et al. 2015*), elements like Co and Cr were found to segregate to SESFs, forming disordered γ phase, which leads to LPT weakening and facilitates the creep microtwinning (*Smith et al. 2015*). However, despite the fact that Ti is an essential constituent of the η phase, its direct role in $D0_{24}$ ordering at SESF, particularly in Co-based and CoNi-based superalloys, remains comparatively unexplored. Moreover, while the involvement of Mo and W, two key γ′-forming elements in Co-based alloys, in promoting $D0_{19}$-type ordering at SISFs has been reported (*Titus et al. 2015*), it remains unclear whether these γ′-phase forming elements can similarly promote local η-phase ordering at SESFs.

In this study, we investigate the creep resistance and related planar faults formation mechanisms in CoNi-based superalloys containing Mo and W, with systematically varied Ti/Al ratios, subjected to creep deformation at 1223 K. We report, for the first time, a transition in the dominant planar faults mode from APB-controlled shearing at low Ti contents to SESF-controlled shearing at high Ti contents. *Ab initio* calculations are employed to evaluate the effects of Ti on APB, CISF, and SESF energetics to explain the transition of dominant planar faults mode, as well as the role of the elements (Mo, W and Ti) in the $L1_2 \rightarrow D0_{24}$ phase transformation. Furthermore, by analyzing solute segregation behaviors, we explain why numerous microtwins form in alloy 8Ti and why local phase transformation does not occur along SESFs, despite the presence of Ti, Mo, and W segregation in the crept sample. Finally, we discussed the relationship between planar faults with varied Ti/Al ratios and creep resistance, providing new insights into how to control Ti/Al ratios to tune the planar faults formation mechanisms to influence high-temperature performance of next-generation CoNi-based superalloys.

## 2. Experiments and Methods

### 2.1 Materials

The alloys under investigation are γ′-phase hardened polycrystalline CoNi-based superalloys named 0Ti, 4Ti and 8Ti, with the nominal chemical composition Co-30Ni-(12.5-x)Al-xTi-2.5Mo-2.5W (x= 0, 4, 8 at.%). The average alloy composition was measured by SEM-EDS and is shown in **Table 1**. The alloys were melted at least 7 times under argon in the form of a 70 g ingot button on a water-cooled copper hearth using a laboratory-scale vacuum arc melting unit. The as-cast material was homogenized at 1523 K for 24 h. In addition, alloys 0Ti, 4Ti and 8Ti with extra 0.1 at.% B were melted again and then homogenized followed by ageing at 1173 K in air for 220 h, and subsequently air cooled to room temperature.

**Table 1. Average alloy composition in atomic percent (at.%) by SEM-EDS.**

| Alloys | Co | ±σ | Ni | ±σ | Al | ±σ | Ti | ±σ | Mo | ±σ | W | ±σ |
|---|---|---|---|---|---|---|---|---|---|---|---|---|
| 0Ti | 53.2 | 0.14 | 30.2 | 0.04 | 11.6 | 0.06 | – | – | 2.4 | 0.04 | 2.6 | 0.03 |
| 4Ti | 52.8 | 0.22 | 30.3 | 0.27 | 7.8 | 0.07 | 4.1 | 0.13 | 2.4 | 0.09 | 2.6 | 0.06 |
| 8Ti | 52.6 | 0.30 | 30.1 | 0.23 | 4.2 | 0.16 | 8.1 | 0.07 | 2.4 | 0.02 | 2.5 | 0.07 |

### 2.2 Creep testing

Compression creep tests of the alloys with extra 0.1% B were performed using a Satec Systems constant load machine with a lever arm ratio of 16:1. The creep tests in alloys 0Ti, 4Ti and 8Ti were performed in air, at 1223 K with compressive stresses of 241 MPa. The creep tests were stopped manually after creep strain of 4.0 %.

### 2.3 Microstructural characterization

The γ/γ′ two-phase microstructures of the aged samples were characterized by backscattered electron (BSE) imaging in a scanning electron microscope (FE-SEM, Zeiss, Germany Gemini) The resultant images of γ'/γ binary phases in alloys 0Ti, 4Ti and 8Ti with extra 0.1 at.% B were shown in **supplementary materials Fig. S1**. A CMOS EBSD Detectors with AZteclive analysis software was used for Electron Backscatter Diffraction (EBSD) measurements. The BSE and EBSD samples were ground using SiC abrasive papers sequentially from 600, 1200 to 4000 grades, followed by vibratory polishing.

The in situ High Energy X-ray diffraction (HEXRD, $\lambda$ = 0.146 Å) for lattice misfit measurement at 1223 K was done at the synchrotron beamline HEMS run by Helmholtz Zentrum Hereon at the PETRA III storage ring of the Deutsches Elektronen-Synchrotron (DESY, Hamburg, Germany). In order to eliminate the coarse grain effect, the cylinder samples were oscillated angularly by ± 40° with three-dimensional XRD scanning.

For transmission electron microscopy (TEM) sample preparation, 3 mm disks were mechanically polished to a thickness of about 75 μm. The final samples were thinned to electron transparency by a twin jet electro polishing unit using the Struers A3 electrolyte with a voltage of 32 V at – 38 °C. High resolution high angle annular dark field (HAADF) imaging was performed in a Thermo Fisher Scientific Themis Z operated at 300 kV with a probe-corrector to investigate the structure of defects in atomic resolution. The Thermo Fisher Scientific Talos 200i equipped with a Super-X EDS detector was operated at 200 kV to analyze the local chemical composition near planar defects. The EDS data were processed using the commercial Velox software. To minimize channeling effects during STEM-EDS acquisition, the sample was intentionally tilted slightly away from the exact zone axis. Quantification was performed using the Cliff-Lorimer k-factor method, and the Cu signal was removed through peak deconvolution. Background removal was carried out using the polynomial background fitting routine in Velox. In this procedure, the Bremsstrahlung background is modeled using a polynomial function, while the characteristic X-ray peaks are fitted using Gaussian profiles. The combined model is iteratively fitted to the raw spectrum, ensuring that the quantified elemental intensities are based on background-subtracted net peak areas.

**2.4 First-principles calculations**

The present investigation focuses mainly on $A_3B$-type ordered structures, i.e., $L1_2$ and $D0_{24}$, with various alloy elements, where Co and Ni occupy A sublattice and other elements (Al, Ti, Mo, and W) reside on B sublattice. Elements on each sublattice are randomly distributed by using the special quasi-random structures (SQS) method (*Zunger et al. 1990*) implemented in the Alloy-Theoretic Automated Toolkit (ATAT) package (*Van de Walle et al. 2013*). The formation energy of the alloy can be determined by the following formula:

$$\Delta H_f = E(\text{alloy}) - \sum c_i E_i, \quad (1)$$

where $E(\text{alloy})$ and $E_i$ are total energies of the alloy and the corresponding constituent element $i$, respectively. $c_i$ is the concentration of element $i$ in the alloy. The $L1_2$ phase was modeled using 32-atom supercells, while 64-atom supercells were used for the $D0_{24}$ phase.

Formation energies of planar defects in the γ' phase were calculated at zero-kelvin with an atomistic model containing 144 atoms. For the $L1_2$ phase, compositions of various elements are designed as $(Co_{37.5}Ni_{37.5})_3(Al_{19.4-m},Ti_m,Mo_3,W_3)$ in which the subscripts represent the composition of elements in the atomistic model. According to previous studies (*Wang et al. 2019, Zhang et al. 2024*), Co and Ni atoms are believed to occupy the A-site while Al, Ti, Mo and W atoms occupy mainly the B-site. In this way, effects of alloying the Ti element were investigated for four systems (alloys 0Ti, 4Ti, 8Ti and 12.5Ti), with $m = 0, 6.25, 12.5, \text{and } 19.44$ corresponding to Ti concentration $c_{\text{Ti}} = 0\ \%, 6.25\ \%, 12{,}50\ \%, \text{and } 19.44\ \%$, respectively.

Planar defects can be sampled in the (111) plane with tilted supercells, as shown in **Fig. 1(a)**. This method has been successfully used to study similar planar defects (*Xu et al. 2023, Xu et al. 2024, Zhang et al. 2018*). Technically, a relative shift between two periodic images of the defect-free structure is implemented by tilting the $\hat{z}$-axis of the supercell by a transition vector $\hat{T}$, as shown by the blue box in **Fig. 1(a)** and **supplementary materials Fig. S2**. Symmetry analysis of a (111) plane identifies 12 symmetry-inequivalent APBs, 12 CISFs, 4 SISFs, and 4 bulk sites, as shown in **Fig. 1(b)**. An SESF is created by introducing a second shift in an SISF model (see **supplementary materials Fig. S2**).

The formation energy of planar defects $\Delta E_{\text{PD}}$ is calculated as:

$$\Delta E_{\text{PD}} = \frac{E_{\text{PD}} - \tilde{E}_{\text{bulk}}}{A_{\text{PD}}},$$

with $E_{\text{PD}}$ the total energy of the planar defect structure. $\tilde{E}_{\text{bulk}}$ is the total energy of the bulk structure averaged over four sites. $A_{\text{PD}}$ is the interface area of a planar defect.

All *ab initio* calculations were performed by using the Projector Augmented Wave (PAW) method (*Blöchl 1990*) implemented in the Vienna Ab-initio Simulation Package (VASP) (*Kresse and Furthmüller 1996*). The exchange-correlation functionals are described by the generalized gradient approximation parameterized by Perdew-Burke-Ernzerhof (PBE) (*Perdew et al. 1996*). All calculations are spin-polarized and with plane-wave cutoff energies of 400 eV or higher. The Methfessel-Paxton broadening scheme (*Methfessel and Paxton 1989*) of the first order was used with a smearing of 0.1 eV. A Γ-centered Monkhorst-Pack grid (*Monkhorst et al.*

*1976*) with a mesh of 4 × 4 × 3 (11×11×11) was used for formation energies of planar defects (defect-free alloys). All supercells were fully relaxed with respect to all degrees of freedom. The convergence criterions of electronic loop and ionic loop for formation energies of defect-free alloys (planar defects) were set to be $10^{-6}$ eV ($10^{-5}$ eV) and $10^{-3}$ eV/Å (0.05 eV/Å).

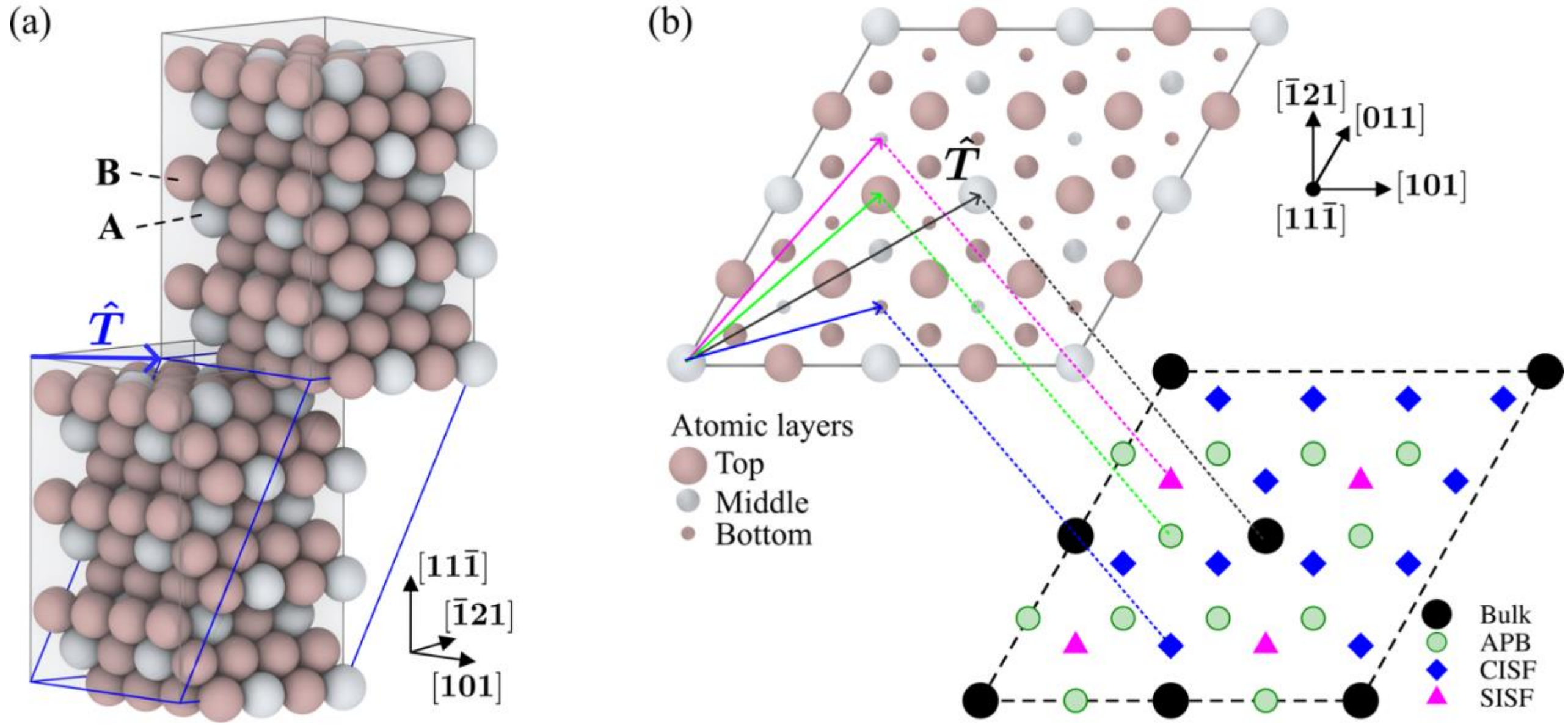

**Fig. 1. Supercell models for sampling planar defects.** (a) An exemplar atomistic model with tilted z-axis. (b) Sites for sampling the planar defects. The symbols in the panel at the right bottom corner represent distinct planar defects when tilting the z-axis to certain positions.

## 3 Results

### 3.1 Creep tests and deformation mechanism analysis by STEM

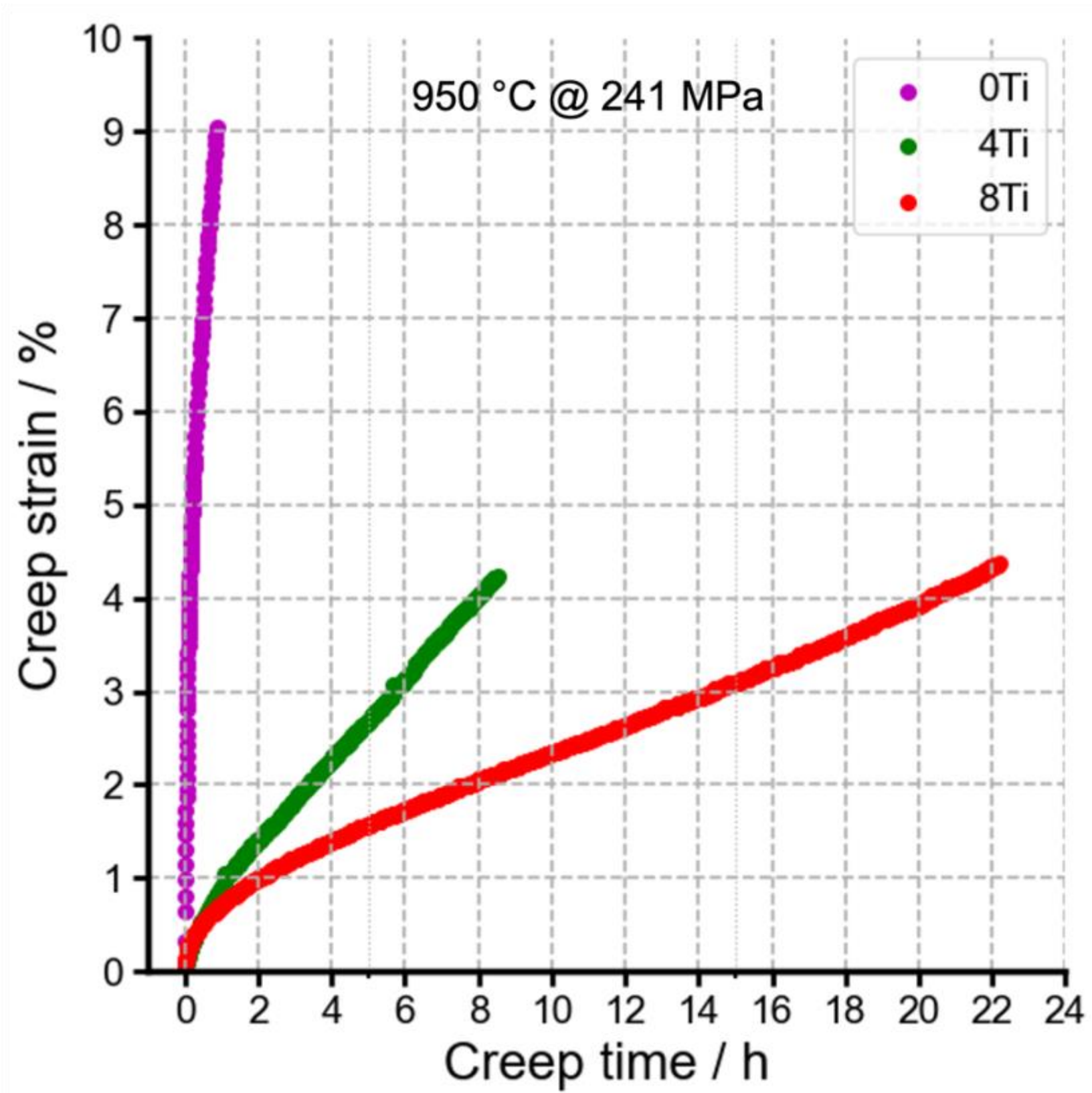

**Fig. 2.** Compression creep test of alloys 0Ti, 4Ti and 8Ti at 1223 K with applied stress of 241 MPa.

**Fig. 2** presents the creep test results of alloys 0Ti, 4Ti and 8Ti at 1223 K with an applied stress of 241 MPa. The high-Ti alloy (8Ti) exhibits much higher creep resistance than the low-Ti (4Ti) and Ti-free (0Ti) alloys. The resultant post-mortem SEM-BSE images for compression creep test of alloys 0Ti, 4Ti and 8Ti are shown in **Fig. 3(a)**. After creep deformation, the γ' particles in alloy 0Ti become irregular while directional coarsening (rafting) appears in alloys 4Ti and 8Ti. Additionally, we found creep microtwins formed in alloy 8Ti. To get a more detailed insight into the microstructure of the crept specimens, the HAADF-STEM (**Fig. 3(b)**) was employed to image defects, including dislocation networks and planar faults (SESF and APBs), near the [011] zone axis. In alloy 0Ti, no dislocation networks were observed. However, a number of planar faults, i.e. APBs, were identified in the γ' precipitates, which suggests that a significant number of dislocations tend to shear the γ' precipitates by glide instead of overcoming them by climb at the γ/γ' interfaces. In contrast, there are dense dislocation networks formed at the γ'/γ interfaces in both alloys 4Ti and 8Ti, however, with Ti content increasing from 4 at.% to 8 at.%, the γ' precipitate shearing mode changed from APBs to SESFs.

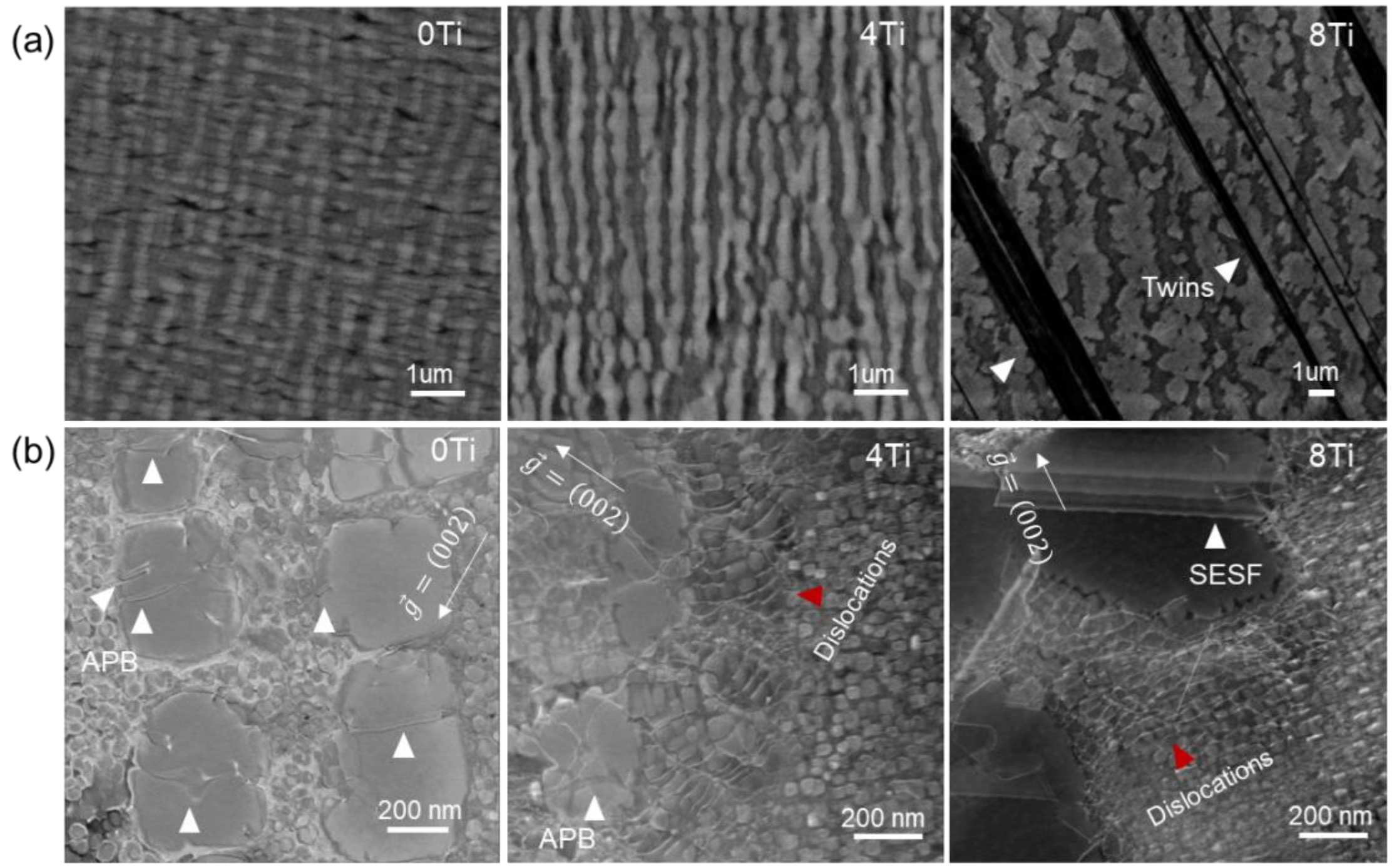

**Fig. 3.** (a) Post-mortem SEM-BSE images for compressive creep specimens of alloys 0Ti, 4Ti and 8Ti. (b) HAADF-STEM (0Ti, 4Ti and 8Ti) images of dislocation networks and planar defects (SESF and APBs) taken near the [110] zone axis. (The white arrows indicate planar defects and red arrows indicate dislocation networks.)

### 3.2 Creep planar defects (microtwins and SESFs) and its chemical fluctuations in alloy 8Ti

To confirm that the structures observed in **Fig. 3(a)** are creep twins in the alloy 8Ti, EBSD measurements were performed. The Kikuchi pattern quality map and the corresponding inverse pole figure (IPF) map are shown in **Fig. 4(a)** and **(b)**. The features highlighted in green exhibit a typical Σ3 twin relationship (60° misorientation) with the surrounding matrix, which is further supported by the misorientation angle distribution in **Fig. 4(c)**. Additional TEM investigations were carried out on a FIB-prepared lamella. The BF-STEM and DF-TEM images in **Fig. 4(d)** and **(e)** reveal the presence of creep microtwins, and the inset diffraction pattern confirms twin reflections. A high-resolution HAADF-STEM image of the microtwin, presented in **supplementary materials Fig. S3**, and its interface presented in **Fig. 4(f)**, where the atomic arrangement across the boundary can be resolved. The vertically integrated intensity line profile across this interface (**Fig. 4(g)**) indicates enrichment of heavy elements, such as Mo and W. Furthermore, chemical fluctuations in the vicinity of planar faults appear to be a key feature during creep deformation of superalloys. The quantified STEM-EDS line scan (**Fig. 4(h)**) reveals segregation of Co, Ti, Mo, and W at the microtwin interface, accompanied by depletion of Ni and Al.

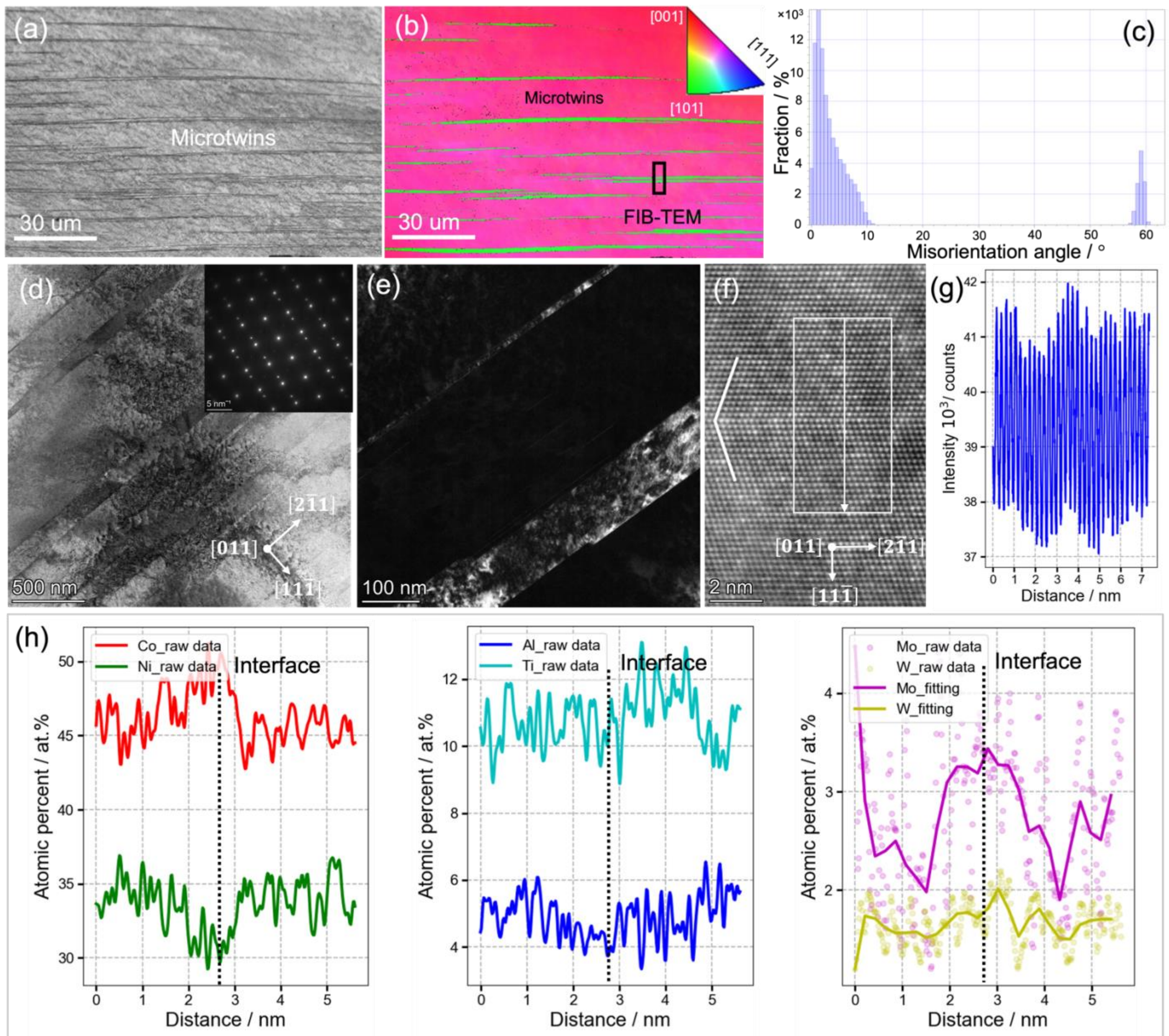

**Fig. 4. Creep microtwinning in alloy 8Ti after creep deformation.** (a) EBSD pattern quality map. (b) Inverse pole figure (IPF) map. (c) Misorientation distribution corresponding to (b). (d) BF-STEM image of a FIB-TEM lamella showing creep microtwins (inset: diffraction pattern confirming twin reflections). (e) DF-TEM image highlighting microtwins. (f) High resolution HAADF-STEM image of a microtwin interface. (g) Vertically

integrated intensity line profile across the microtwin interface. (h) STEM-EDS compositional line profile across the microtwin interface.

The majority of the observed shearing events in the alloy 8Ti were identified as isolated SFs within the $\gamma'$ precipitates, as shown in **Fig. 5(a)** and **(b)**. According to the established criteria for distinguishing between extrinsic and intrinsic SFs (**supplementary materials Fig. S4**), the direction of the g-vector points away from the outermost bright fringe of the stacking fault. On this basis, the most of isolated SFs were classified as extrinsic SFs.

Atomic-resolution HAADF-STEM images of two SESFs, labeled SF1 and SF2, are presented in **Fig. 5(c)** and **(d)**. **Fig. 5(c)** shows the trailing part of both SESFs, while **Fig. 5(d)** displays their leading part. All Shockley partial dislocations are aligned parallel to the line direction, $\boldsymbol{\mu}$ = [011], which corresponds to the **DB** direction in the tetrahedron illustrated in **Fig. 5(e)**. The Burgers vector analysis was carried out following the conventional criterion, where the vector is defined from the finish point (F) to the start point (S) of the Burgers circuit. Clockwise Burgers circuits were drawn around the leading and trailing edges.

The trailing partial of the two SESFs in crystal plane $(11\bar{1})$, shown in **Fig. 5(c)**, exhibits two Burgers circuits with two projected Burgers vector of $\mathbf{b}_{1,p}$ = $\mathbf{F_1S_1}$ = $a/6[\bar{2}1\bar{1}]$ and $\mathbf{b}_{2,p}$ = $\mathbf{F_2S_2}$ = $a/6[\bar{2}1\bar{1}]$ in the projection plane. The corresponding real Burgers vector are deduced to be $\mathbf{b}_1$ = $a/3[\bar{1}21]$ (2**γB**) and $\mathbf{b}_2$ = $a/3[\bar{1}21]$ (2**γB**). Given the 30° angle between the dislocation line and Burgers vector, the trailing partial **γB** is identified as a 30° mixed Shockley partial. Actually, $\mathbf{b}_1$ = $\mathbf{b}_2$ = 2**γB** means each trailing segment consists of two identical 30° mixed Shockley partials.

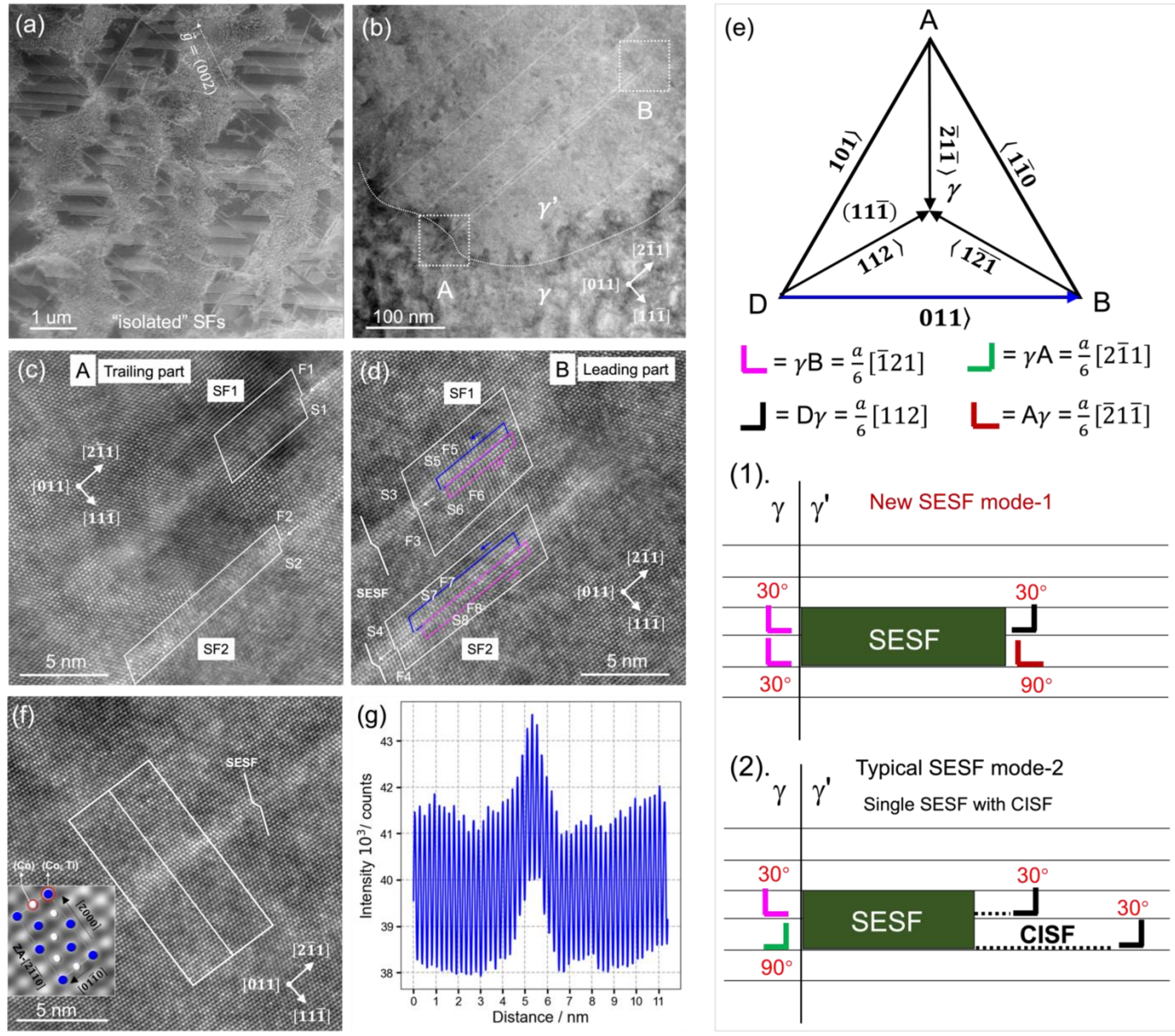

**Fig. 5. Structural characterization of isolated SFs and SESFs in alloy 8Ti.** (a) HAADF-STEM image of isolated SFs near the [011] zone axis with **g** = (002). (b) HAADF-STEM image of isolated SFs along the [011] zone axis. (c) and (d) High-resolution HAADF-STEM images of two SESFs (**SF1** and **SF2**), with Burgers circuits used to determine their configurations. (e) Schematic model of the new SESF mode-1 (1) proposed in this work, compared with the typical SESF mode-2 (2) reported in literatures (*Kolbe 2001, Smith et al. 2016, Kovarik et al. 2009, Smith et al. 2015, Karpstein et al. 2023, Lilensten et al. 2021*). (f) and (g) HAADF high-resolution HAADF-STEM image of an SESF (Inset: $D0_{24}$ atomic structure at the faulted region) and its integrated intensity line profile which indicates there is strong heavy elements segregation along SESFs.

For the leading partial dislocations (LPDs) of two SESFs in crystal plane $(11\bar{1})$, shown in **Fig. 5(d)**, six Burgers circuits were analysed. The corresponding projected Burgers vectors are: $\mathbf{b}_{3,p}$ = $\mathbf{F}_3\mathbf{S}_3$ = $a/12[\bar{2}1\bar{1}]$ $\mathbf{b}_{5,p}$ = $\mathbf{F}_5\mathbf{S}_5$ = $a/6[\bar{2}1\bar{1}]$ and $\mathbf{b}_{6,p}$ = $\mathbf{F}_6\mathbf{S}_6$ = $a/12[2\bar{1}1]$ in **SF1** and $\mathbf{b}_{4,p}$ = $\mathbf{F}_4\mathbf{S}_4$ = $a/12[\bar{2}1\bar{1}]$, $\mathbf{b}_{7,p}$ = $\mathbf{F}_7\mathbf{S}_7$ = $a/6[\bar{2}1\bar{1}]$ and $\mathbf{b}_{8,p}$ = $\mathbf{F}_8\mathbf{S}_8$ = $a/12[2\bar{1}1]$ in **SF2**, all within the projection plane. The corresponding true Burgers vectors are determined as: $\mathbf{b}_3$ = $a/6[\bar{1}21]$ (denoted **γB**), $\mathbf{b}_5$ = $a/6[\bar{2}1\bar{1}]$ (**A**γ) and $\mathbf{b}_6$ = $a/6[112]$ (**D**γ) in **SF1** and $\mathbf{b}_4$ = $a/6[112]$ (γ**B**), $\mathbf{b}_7$ = $a/6[\bar{2}1\bar{1}]$ (**A**γ) and $\mathbf{b}_8$ = $a/6[112]$ (**D**γ) in **SF2**. Here, **γB** is identified as a 30° mixed Shockley partial**,** and **A**γ as a 90° edge Shockley partial**.** Notably, the vector relations $\mathbf{b}_3 = \mathbf{b}_5 + \mathbf{b}_6$ and $\mathbf{b}_4 = \mathbf{b}_7 + \mathbf{b}_8$ confirm that the LPDs in both SESFs consists of one 90° edge Shockley partial (**A**γ) and one 30° mixed Shockley partial (**D**γ). Further explanations for distinguishing between 30° mixed-type and 90° edge-type Shockley partials, as well as the method for determining the real Burgers vectors from the projected ones in HRSTEM images, are provided in **supplementary materials Fig. S4**. This analysis demonstrates that the two SESFs share the same dislocation configuration, as schematically illustrated in **Fig. 5(e)-(1)**. Accordingly, the dislocation reactions in **SF1** and **SF2** can be expressed as:

$$\begin{cases} \underset{\frac{a}{2}[011]}{\mathbf{0°, DB}} \rightarrow \underset{\frac{a}{6}[112]}{\mathbf{30°, D\gamma}}(\text{leading}) + \mathbf{CISF} + \underset{\frac{a}{6}[\bar{1}21]}{\mathbf{30°, \gamma B}}(\text{trailing}) \\ \underset{\frac{a}{2}[\bar{1}10]}{\mathbf{60°, AB}} \rightarrow \underset{\frac{a}{6}[\bar{2}1\bar{1}]}{\mathbf{90°, A\gamma}}(\text{leading}) + \mathbf{CISF} + \underset{\frac{a}{6}[\bar{1}21]}{\mathbf{30°, \gamma B}}(\text{trailing}) \end{cases} \tag{1}$$

The SESF configuration observed here represents a new defect mode distinct from those previously reported. In the majority of studies, SESFs are associated with the typical mode containing a single CISF (*Kolbe 2001, Smith et al. 2016, Kovarik et al. 2009, Smith et al. 2015, Karpstein et al. 2023, Borovikov et al. 2023, Borovikov et al. 2024*), as illustrated in **Fig. 5(e)-(2).** The SESF mode-2 was reported by Karpstein et al. (*Karpstein et al. 2023*), who experimentally demonstrated that the two dislocations bounding a CISF correspond to identical Shockley partials. The dislocation reactions can be expressed as below:

$$\begin{cases} \underset{\frac{a}{2}[011]}{\mathbf{0°, DB}} \rightarrow \underset{\frac{a}{6}[112]}{\mathbf{30°, D\gamma}}(\text{leading}) + \mathbf{CISF} + \underset{\frac{a}{6}[\bar{1}21]}{\mathbf{30°, \gamma B}}(\text{trailing}) \\ \underset{\frac{a}{2}[101]}{\mathbf{60°, DA}} \rightarrow \underset{\frac{a}{6}[112]}{\mathbf{30°, D\gamma}}(\text{leading}) + \mathbf{CISF} + \underset{\frac{a}{6}[2\bar{1}1]}{\mathbf{90°, \gamma A}}(\text{trailing}) \end{cases} \tag{2}$$

In the newly proposed SESF mode-1, the angle between the Shockley partials **Dγ** and **Aγ** is 120° which causes the Shockley partials to attract and combine into a **γB** dislocation. As a result, the corresponding CISF is not clearly visible. By contrast, in SESF mode-2, the two Shockley partials (**Dγ**) possess the same sign, leading to mutual repulsion. This produces a single CISF bounded by the two Shockley partials, making the complex stacking fault clearly observable (*Karpstein et al. 2023, Vorontsov et al. 2012*). These observations are consistent with the Kolbe mechanism (*Kolbe 2001*), in which two $a/2\langle 110\rangle$ dislocations within the γ matrix are disassociated into produce four Shockley partial dislocations. Under deformation at high temperature, the movement of leading Shockley partials are activated and are capable of shearing the $\gamma'$ precipitate. Shearing of two leading Shockley partials within the $\gamma'$ phase can result in the formation of a two-layer CISF, where the nearest-neighbor violations occur in its wake, requiring local atomic re-ordering to establish a true SESF via segregation of alloying elements, reducing its high-energy penalty and thereby transforming it into a more stable, lower-energy condition.

The enhanced Z-contrast observed in the fault regions of **Fig. 5(f)** is consistent with local structural changes. Nevertheless, there is no grid-like ordering present along the SESF based on Z-contrast analysis in HAADF-STEM image. The $D0_{24}$ simulation used here is based on the $Co_3Ti$ structure, in which blue columns represent Co+Ti atoms and gray columns represent Co atoms.  The HAADF integrated intensity line profile in **Fig. 5(g)** indicates there is strong elements segregation of heavy elements, such as Mo and W, compared to the surrounding $\gamma'$ lattice.

To confirm this, STEM-EDS was used to analyze the local chemistry. A low-magnification HAADF-STEM image (**Fig. 6(a)**) reveals several SESFs formed by shearing in a $\gamma'$ precipitate. Elemental maps (**Fig. 6(b)**) show clear segregation of Co, Mo, and W along the SESFs, accompanied by depletion of Ni and Al. The integrated line scan (**Fig. 6(c)**) quantitatively confirms enrichment of Co, Mo, W, and to a lesser extent Ti, as well as depletion of Ni and Al. In addition, a Co-rich Cottrell atmosphere is observed around the leading Shockley partial dislocations (**supplementary materials Fig. S5**). The *ab initio* predictions presented in **Section 4.2** further prove whether high Ti, W, and Mo additions could make SESF toward $D0_{24}$ structure transformation.

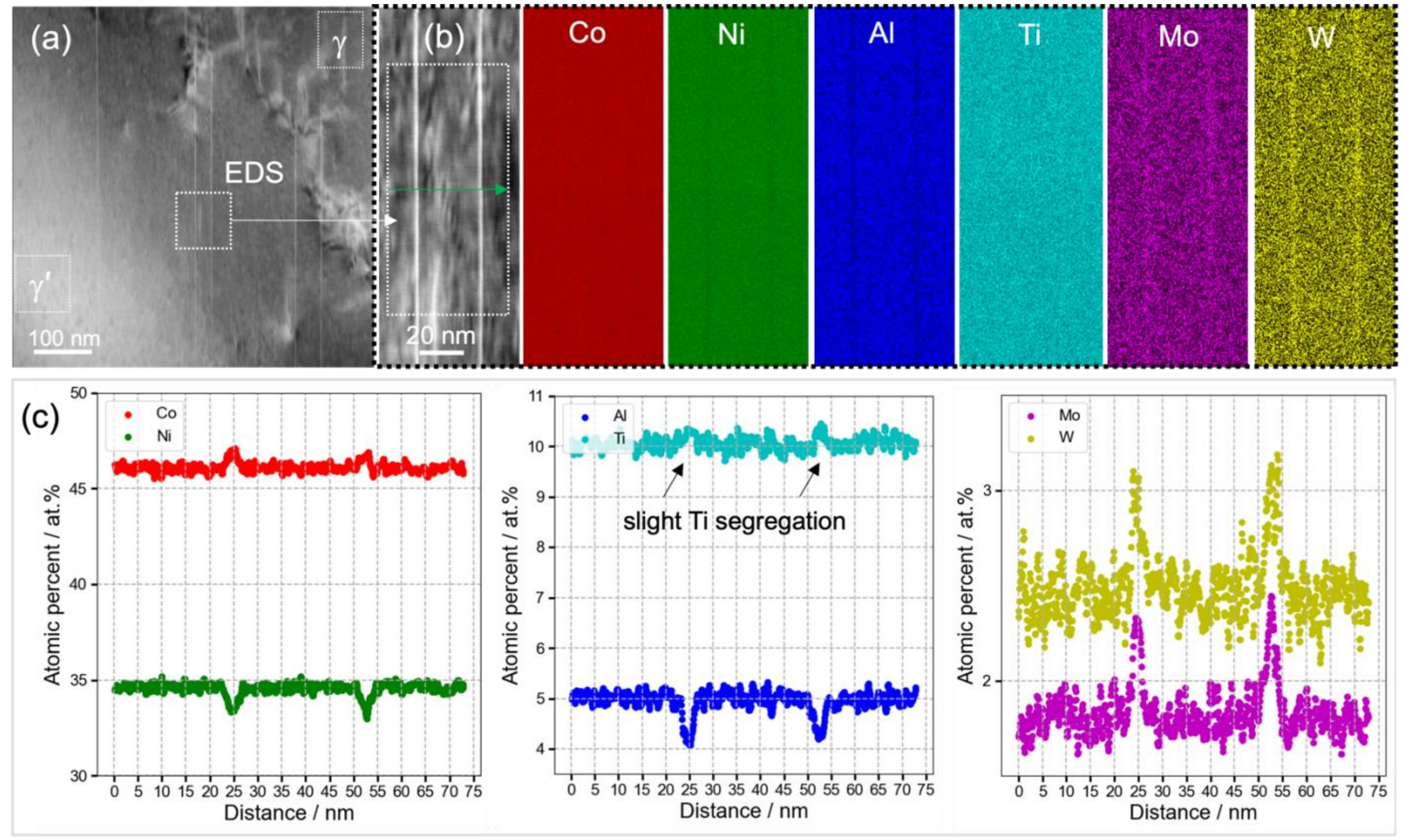

**Fig. 6. Chemical fluctuations analysis of isolated SESFs in alloy 8Ti.** (a) HAADF-STEM image of SESFs in [011] beam direction. (b) Net intensity elemental maps of two vertical SESFs. (c) The integrated EDS line scanning curves represent the area incorporated into the vertically integrated line scan shown from **(b)**.

### 3.3 Chemical fluctuations at planar faults of APB in alloys 0Ti and 4Ti

**Fig. 7** shows a HAADF-STEM image and corresponding elemental maps of an $APB_{(111)}$ on the (111) plane inside a γ' precipitate in the alloy 0Ti. The elemental maps (**Fig. 7(d)-(h)**) reveal pronounced segregation of Co at the $APB_{(111)}$, accompanied by depletion of Ni, Al, Mo, and W. An integrated line scan across the region of interest, shown in **Fig. 7(i)** and **(j)**, quantitatively confirms this compositional redistribution. In addition, the EDS maps indicate that Co enrichment and Ni, Al, Mo, and W depletion also occur at the leading dislocation terminating the APB within the γ' precipitate, similar to the chemistry observed at the APB itself.

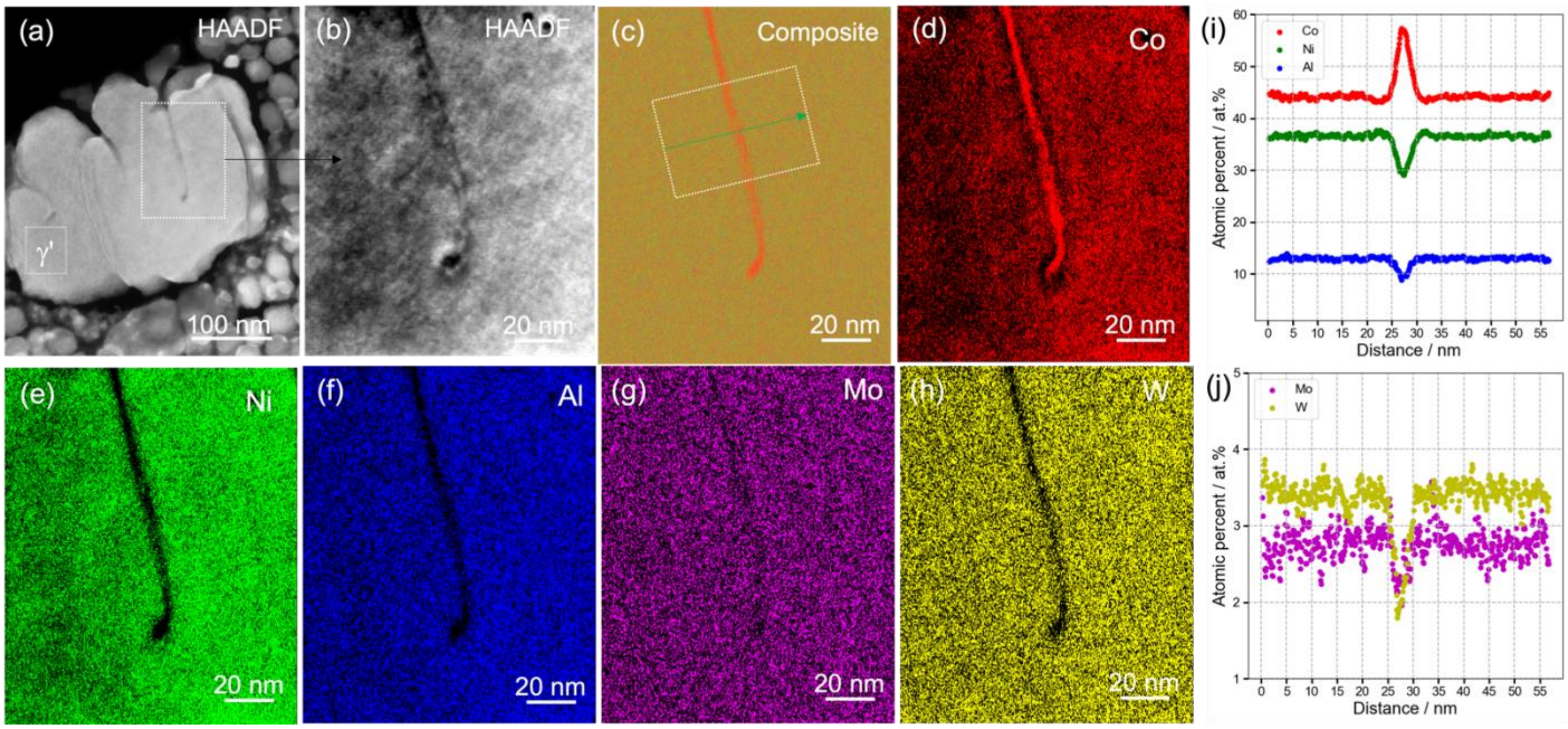

**Fig. 7. Chemical fluctuations analysis around an APB region on a (111) plane in alloy 0Ti.** (a) HAADF-STEM image of the γ' precipitate with APBs taken along [011] beam direction. (b) Magnified image of white rectangular marked in **Fig. 7(a)**. (c) Composite chemical map of elements Co, Ni, Al, Mo and W. (d)-(h) Net intensity elemental maps of elements Co, Ni, Al, Mo and W. (i) and (j) EDS line scan integrated along the APB in the region marked in **Fig. 7(c)**.

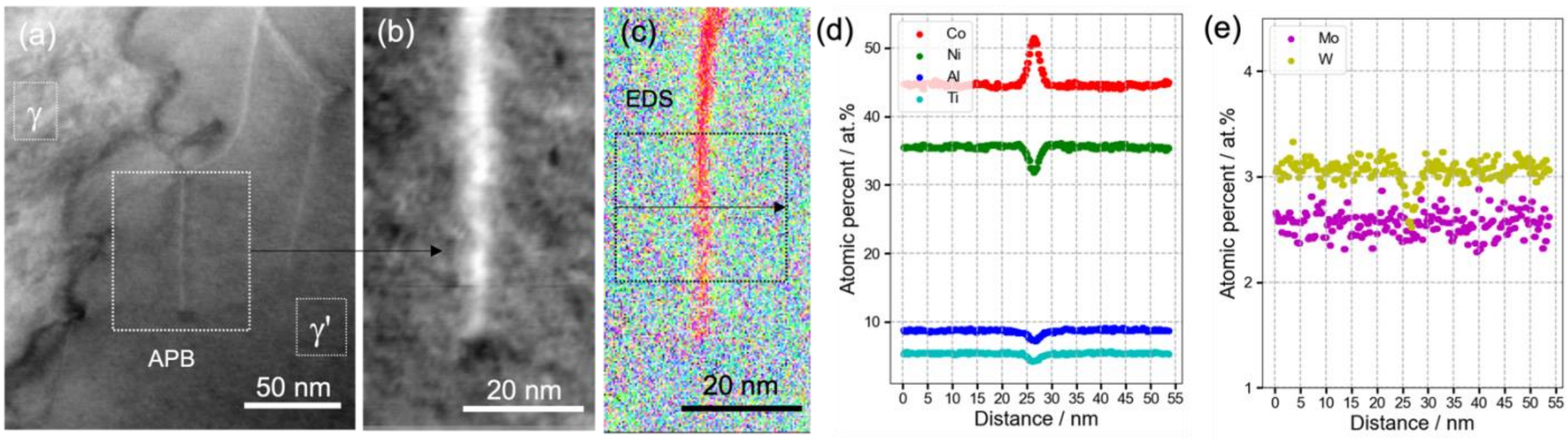

**Fig. 8. Chemical fluctuations analysis around an APB region on a (001) plane in alloy 4Ti.** (a) HAADF-STEM image of the γ' precipitate with an APB taken along [001] beam direction. (b) Magnified image of white rectangular marked in **Fig. 8(a)**. (c) Composite chemical map of elements Co, Ni, Al, Ti, Mo and W. (d) and (e) EDS line scan integrated along the APB in the region marked in **Fig. 8(c)**.

**Fig. 8** presents HAADF-STEM images (**Fig. 8(a), (b)** and a composite chemical map **Fig. 8(c)**) of a region containing an $APB_{(001)}$ on the (001) plane inside a γ' precipitate in the alloy 4Ti. Each chemical map of elements Co, Ni, Al, Mo and W was shown in **supplementary materials Fig. S7**. An integrated line scan across this region, shown in **Fig. 8(d)** and **(e)**, reveals pronounced Co segregation along the $APB_{(001)}$, accompanied by depletion of Ni, Al, Ti, and W. Similar segregation behavior was observed for APB on (111) planes (**supplementary materials Fig. S8**), confirming that APBs on both (001) and (111) planes correspond to localized regions of disordered γ phase.

The partitioning of alloying elements between the γ and γ' phases in the 0Ti, 4Ti, and alloy 8Tis was further analyzed by STEM-EDS (**supplementary materials Fig. S9**). In the alloy 0Ti, only Co preferentially partitions into the γ matrix, whereas Ni, Al, Mo, and W partition into the γ' phase. In the alloy 4Ti, Co again partitions into the γ matrix, while Ni, Al, Ti, and W partition into the γ' phase; Mo, however, exhibits a nearly balanced distribution between the two phases. In the alloy 8Ti, both Co and Mo partition preferentially into the γ matrix, while Ni, Al, Ti, and W partition into the γ' phase. With increasing Ti content, Mo changes its partitioning behavior from a γ'-former (0Ti) to a γ-former (8Ti).

These results highlight that the element partitioning tendencies in the γ/γ' two-phase microstructure are consistent with the segregation trends observed at APBs within the γ' precipitates. Consequently, APBs inside the γ' phase can be interpreted as localized regions of disordered γ phase.

### 3.4. Solute segregation driven local phase transformations at planar defects

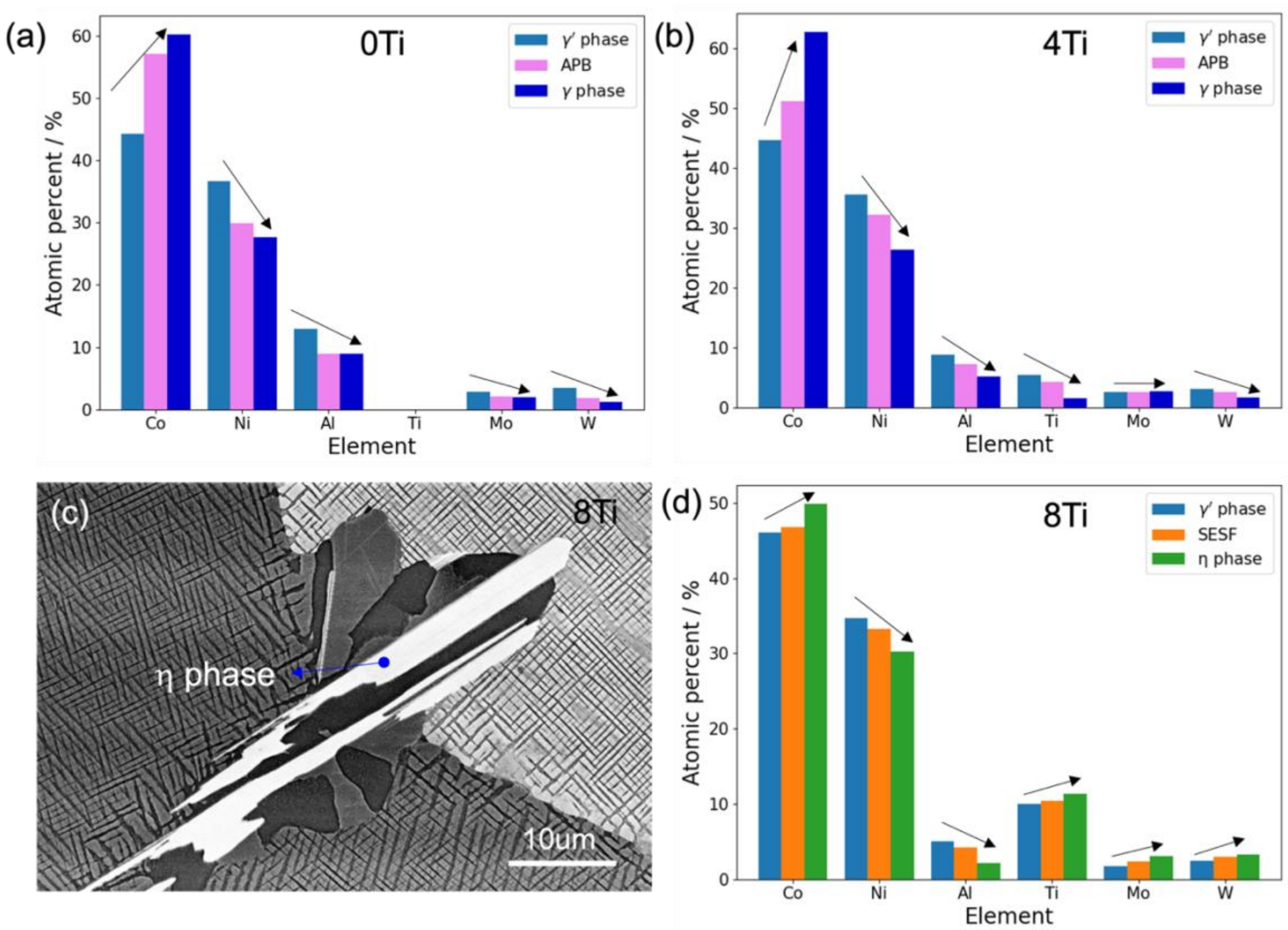

**Fig. 9.** (a) and (b) Comparison of the compositions (at.%) of the γ' phase, APB region and γ matrix in alloys 0Ti and 4Ti. (c) SEM-BSE image with the coarse lath-like η phase in alloy 8Ti after 1036 h aging heat treatment at 900 °C. (d) Comparison of the compositions (at.%) of the γ' phase, SESF region and lath η phase in alloy 8Ti. (The composition details were shown in **Table 2**.)

**Fig. 9(a) and (b)** compares the atomic compositions (at.%) of the γ′ phase, APB regions, and γ matrix in the alloys 0Ti and 4Ti, respectively. The comparison reveals that segregation around APBs promotes LPT from the ordered γ′ phase to the disordered γ matrix. In both the alloys 0Ti and 4Ti, the primary γ-matrix forming element is Co. Several studies (*Eggeler et al. 2016, He et al. 2020, Lu et al. 2020, Borovikov et al. 2025*) have reported on the structural disorder of APBs on (111) planes in the γ' phase using high-resolution STEM. Insights into the local chemistry of planar faults have been further advanced by aberration-corrected STEM combined with high-sensitivity EDS. Yolita et al. (*Eggeler et al. 2016*) observed residual ordering contrast along the full length of the APB on (111) planes, suggesting that the broadened APB retains a degree of $L1_2$ atomic ordering, even though its chemical composition approaches that of the disordered γ phase. In contrast, He et al. (*He et al. 2020*) and Lu et al. (*Lu et al. 2020*) analyzed fast Fourier transform (FFT) patterns along the [110] zone axis from APB regions on (111) planes and reported the disappearance of superlattice diffraction spots, indicating that these APBs had transformed into a disordered γ structure. We also observed the residual ordering along the APB domain on (001) planes in alloy 4Ti by HRTEM, as shown in **supplementary materials Fig. S10**.

The alloy 8Ti was subjected to long-term aging at 900 °C for 1036 h, the bright lath-like η phases appeared along grain boundaries, as shown in **Fig. 9(c)**. The composition of the η laths was measured by SEM-EDS and compared with that of the γ′ phase, SESFs, and bulk η phase in the alloy 8Ti (**Fig. 9(d), Table 2**). Relative to γ′, both the η laths and the SESFs exhibit enrichment in Co, Ti, Mo, and W, accompanied by depletion in Ni and Al.

Although the equilibrium composition of SESFs within the γ′ precipitates does not match that of the bulk $(Co,Ni)_3(Ti,Mo,W)$-$D0_{24}$ phase of the same crystal structure, a clear compositional trend is observed: from γ′ → SESFs → bulk η, the concentrations of Co, Ti, Mo, and W progressively increase, whereas Ni and Al concentrations decrease. In the alloy 8Ti, the η-phase formers are Co, Ti, Mo, and W, with Ti, Mo, and W sharing the same sublattice positions. The segregation of Ti, Mo and W at SESFs is proven to have local η phase transformation possibility, which is further discussed by *ab initio* calculations in **section 4.2**.

**Table 2.** Compositions (at.%) of the γ' phase, APB and γ matrix measured by STEM-EDS in alloys 0Ti and 4Ti and SESF measured by STEM-EDS and bulk η phase measured by SEM-EDS in alloy 8Ti, respectively.

| Alloys | Elements | Co | Ni | Al | Ti | Mo | W |
|---|---|---|---|---|---|---|---|
| **0Ti** | **γ' phase** | 44.3 | 36.6 | 12.9 | - | 2.8 | 3.4 |
| | **APB** | 57.1 | 29.9 | 8.9 | - | 2.1 | 1.8 |
| | **γ matrix** | 60.2 | 27.7 | 8.9 | - | 1.9 | 1.2 |
| **4Ti** | **γ' phase** | 44.7 | 35.5 | 8.8 | 5.4 | 2.6 | 3.1 |
| | **APB** | 51.1 | 32.2 | 7.3 | 4.3 | 2.6 | 2.5 |
| | **γ matrix** | 62.7 | 26.3 | 5.1 | 1.5 | 2.7 | 1.6 |
| **8Ti** | **γ' phase** | 46.10 | 34.64 | 5.01 | 10.00 | 1.75 | 2.49 |
| | **SESF** | 46.82 | 33.26 | 4.22 | 10.36 | 2.36 | 2.97 |
| | **Bulk η phase** | 49.86 | 30.26 | 2.17 | 11.38 | 3.02 | 3.31 |

## 4 Discussion

In the present study, we demonstrated that different types of planar faults are formed during creep deformation depending on the Ti content. Specifically, APBs are predominant at low Ti levels, whereas SESFs appear at higher Ti concentrations in CoNi-based superalloys. To elucidate this behavior, the variations in the calculated APB, CISF, and SESF energies with Ti concentration in the $L1_2$ structure are analyzed to rationalize the formation of specific fault types during deformation. In addition, the elements segregation along SESFs and microtwins observed in high-Ti CoNi-based superalloys is discussed in detail. Finally, the influence of planar faults on the creep resistance is evaluated and the role of the volume fraction of γ′ precipitates and the γ/γ′ lattice misfit in creep resistance of alloy 8Ti is discussed.

**4.1 Planar faults and their formation energies**

Shearing mechanisms of γ' precipitates in Co-, CoNi-, and Ni-based superalloys have been extensively studied over a wide range of temperatures (*Lenz et al. 2019, Barba et al. 2017, Kear et al. 1968, Kolbe 2001, Smith et al. 2016, Kovarik et al. 2009, Smith et al. 2015, Karpstein et al. 2023, Feng et al. 2018, Titus et al. 2015*). The formation and evolution of planar defects during shearing are strongly influenced by both temperature and alloy composition. At temperatures above ~ 850 °C, shearing of the γ' phase in Ni-based superalloys is primarily dominated by APB coupling of $a/2\langle 110\rangle$ superlattice dislocations (*Reed 2008, Feng et al. 2018*). In contrast, in Co-based superalloys such as Co-9Al-9W, γ' precipitate shearing occurs mainly via the formation of SISFs, associated with a leading $a/3\langle 112\rangle$ and a trailing $a/6\langle 112\rangle$ dislocation (*Titus et al. 2015*). Titus et al. (*Titus et al. 2015*) reported that the predominant planar fault type transforms from SISFs to APBs with increasing Ni concentration, which was attributed to the difference between SISF and APB formation energies. Moreover, Ti additions in γ'-strengthened superalloys, Ni-, CoNi-, and Co-based alike, are known to increase the APB energy (*Mottura et al. 2012, Titus et al. 2016, Rhein et al. 2018, Chandran et al., 2011, Wang et al. 2018*) by enhancing $L1_2$ structure ordering, which typically improves high-temperature strength by impeding dislocation motion. However, the effect of Ti on other planar faults, especially CISFs and SESFs, has not been systematically studied.

In the present work, we observed a composition-dependent transition of planar fault types, from APBs to SESFs, with increasing Ti concentration in CoNi-based superalloys. Upon $a/2\langle 110\rangle$ shearing of γ′ precipitates, the high-energy fault states imposed by CISFs and APBs, both involving nearest-neighbor violations, are initially formed. Subsequent chemical reordering lowers the fault energy, thereby promoting transformation from high-energy faults (e.g., CISFs) to lower-energy ones (e.g., SESFs). Therefore, a direct energetic criterion is to compare the variation trend of CISF, APB, and SESF formation energies to determine the preferred fault type under different alloy chemistries.

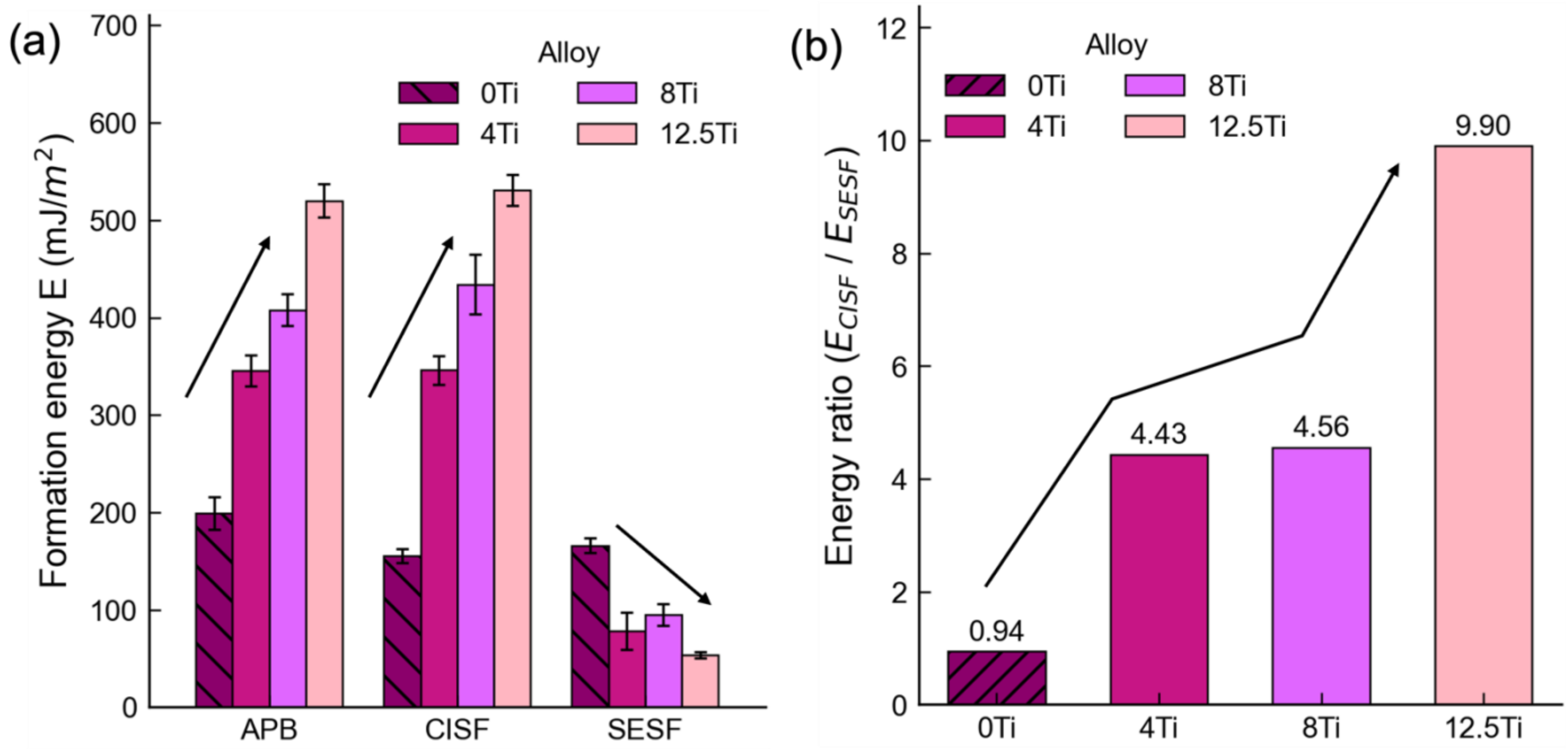

**Fig. 10.** (a) $E_{\mathrm{APB}}^{(111)}$, $E_{\mathrm{CISF}}^{(111)}$ and $E_{\mathrm{SESF}}^{(111)}$ energies (mJ/m$^2$) calculated by the *ab initio* calculations for the $L1_2$ phases with compositions corresponding to the γ' phase in alloys 0Ti, 4Ti, 8Ti, and 12.5Ti: $L1_2$-$(Co_{37.5},Ni_{37.5})_3(Al_{19.44},Mo_3,W_3)$ (0Ti), $L1_2$-$(Co_{37.5},Ni_{37.5})_3(Al_{13.19},Ti_{6.25},Mo_3,W_3)$ (4Ti), $L1_2$-$(Co_{37.5},Ni_{37.5})_3(Al_{6.94},Ti_{12.50},Mo_3,W_3)$ (8Ti) and $L1_2$-$(Co_{37.5},Ni_{37.5})_3(Ti_{19.44},Mo_3,W_3)$ (12.5Ti). The supercell structures were constructed according to the experimentally measured γ' compositions obtained by STEM-EDS (*Liang et al. 2024*). (b) The ratio of $E_{\mathrm{CISF}}^{(111)}$ and $E_{\mathrm{SESF}}^{(111)}$ (mJ/m$^2$) for the same compositions, based on the calculated results shown in **(a)**.

Systematic *ab initio* calculations have been performed for the APB, CISF, and SESF formation energies in the γ' phase, shown in **Fig. 10(a)** and **supplementary materials Table S2**. Both the APB and CISF energies increase monotonically with increasing Ti concentration, whereas the SESF energy decreases on average. The increasing APB energies account for the progressive suppression of APB shearing, leading to the absence of APBs in CoNi-based superalloys containing high Ti additions. Consistent with a gradual suppression with increasing Ti, fewer APBs were observed in the crept microstructures of low-Ti alloys compared with Ti-free ones (**Fig. 3**). The opposite Ti-dependence of APB and SESF energies rationalizes the reduced occurrence of APBs and the dominance of SESFs in high-Ti CoNi-based alloys, consistent with the microstructural observations presented in **Fig. 3**.

Among all investigated compositions, the SESF energy is the lowest in the Al-free alloy (12.5Ti). The corresponding value is 53.5 mJ·m⁻², significantly lower than that in the alloy 8Ti (95.1 mJ·m⁻²). Consistent with this strong compositional dependence, slight Ti segregation is observed along SESFs in the alloy 8Ti, locally enriching the faulted region toward a composition similar to that of the alloy 12.5Ti. This observation suggests that the Ti segregation along SESFs in alloy 8Ti acts as a local energy-reduction mechanism. This stabilization mechanism is also consistent with prior reports that low-energy SESFs can evolve from high-energy CISFs through a chemical reordering process (*Kolbe 2001, Smith et al. 2016, Kovarik et al. 2009, Smith et al. 2015, Karpstein et al. 2023, Lilensten et al. 2021*). To evaluate this transformation tendency, the ratio $E_{\mathrm{CISF}}/E_{\mathrm{SESF}}$ was calculated for different Ti concentrations, as shown in **Fig. 10(b)**. The ratio increases with Ti content, indicating an increasingly strong energetic bias toward CISF → SESF transformation in high-Ti alloys.

A schematic illustration of γ' shearing mechanisms during high-temperature creep in Ti-free, low-Ti, and high-Ti superalloys is presented in **Fig. 11**. In low-Ti alloys, Co segregation (a γ-phase former) and depletion of γ' formers (Ni, Al) reduce the stability of APBs, softening the fault and facilitating creep. In high-Ti alloys, however, SESFs dominate as the main planar defect in the γ' phase. Unlike APBs, SESFs exhibit strong segregation of Co, Ti, Mo, and W, forming a local composition towards $(Co,Ni)_3(Ti,Mo,W)$ to stabilize the SESF.

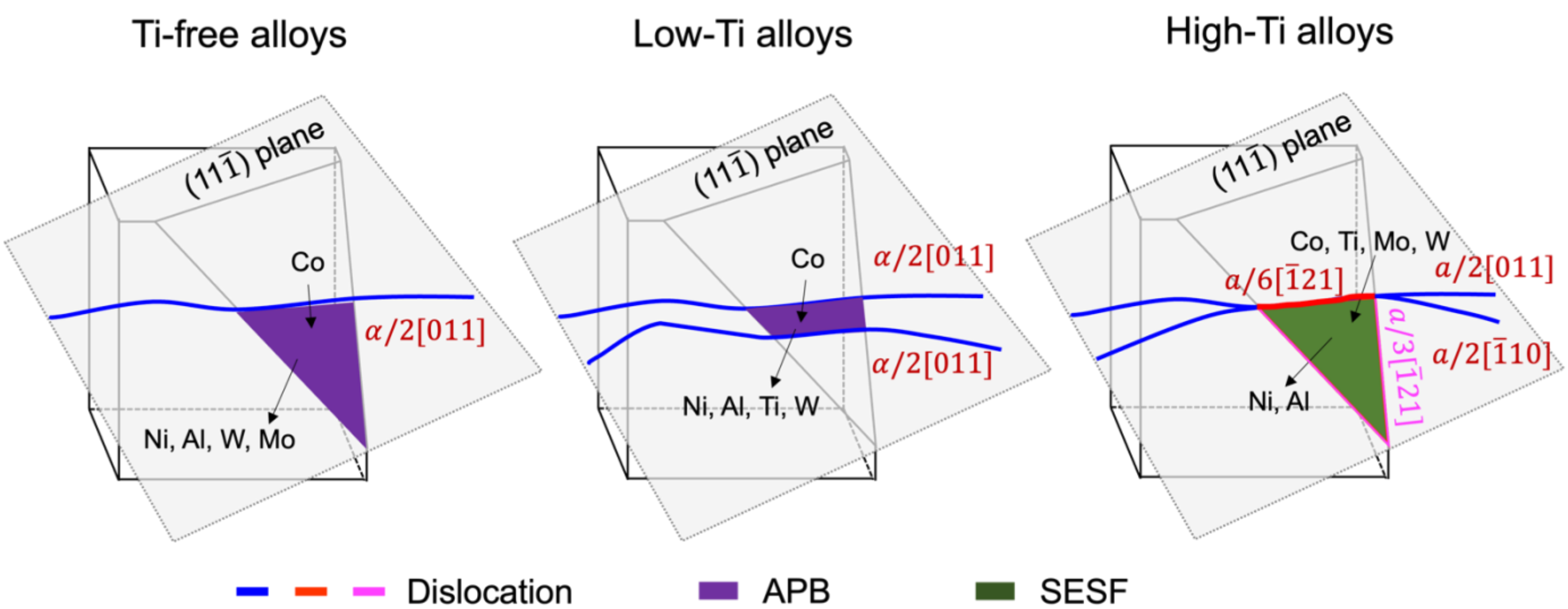

**Fig. 11. Schematic illustrations of dislocation-precipitate shearing mechanisms during creep at high temperatures, i.e.** 1223 K**, in Ti-free, low-Ti and high-Ti CoNi based superalloys.** In Ti-free alloys, the low APB energy allows a single $a/2[011]$ perfect dislocation (blue line) to sweep through the γ′ precipitates. In

low-Ti alloys, the increased APB energy resulting from enhanced $L1_2$ ordering requires a pair of $a/2[011]$ perfect dislocations (blue line) in $(11\bar{1})$ plane to shear the $\gamma'$ phase; the separation between these dislocations remains small due to the high APB energy. In high-Ti alloys, *ab initio* calculations reveal that the APB energy becomes sufficiently high to suppress APB formation entirely. Instead, shearing occurs through the activation of SESFs consisting of two leading Shockley partials ($\mathbf{D\gamma} + \mathbf{A\gamma} = a/6\,[\bar{1}21]$, red line) and two trailing Shockley partial ($\mathbf{2\gamma B} = a/3\,[\bar{1}21]$, magenta line) in $(11\bar{1})$ plane. Elemental segregation along the SESFs stabilizes the faulted region, promoting local structural transition toward the low-energy state.

**4.2 The role of Ti, Mo and W in the η phase transformation**

The concept of reordering, i.e., chemical segregation, has been proposed as a local, diffusion-mediated process occurring around faulted structures. In Co-based superalloys, such segregation occurs preferentially along SISFs. These nanoscale SISFs with W segregation share the same crystal structure as the equilibrium $Co_3W$ ($D0_{19}$) laths and, through segregation, locally attain a composition closer to that of bulk $Co_3W$ (*Titus et al. 2016*). However, the influence of Mo and W on LPT from SESFs to the η ($D0_{24}$) phase remains unclear in CoNi-based superalloys.

In this work (**Fig. 9**), we observed segregation of Co, Ti, Mo, and W along SESFs, shifting the local composition toward that of the η phase. The *ab initio* calculations of the formation energies in **Fig. 12** for $(Co_{37.5},Ni_{37.5})_3(Al_{25-x},Ti_x)$, $(Co_{37.5},Ni_{37.5})_3(Al_{22-x},Ti_x,Mo_3)$, $(Co_{37.5},Ni_{37.5})_3(Al_{22-x},Ti_x,W_3)$ and $(Co_{37.5},Ni_{37.5})_3(Al_{19-x},Ti_x,Mo_3,W_3)$ indicate that at low Ti contents, the $L1_2$-structured $(Co,Ni)_3(Al,Ti)$ is more stable than the $D0_{24}$ structure, whereas at higher Ti contents, the $D0_{24}$-structured $(Co,Ni)_3(Al,Ti)$ phase becomes energetically favorable.

For the $(Co,Ni)_3(Al,Ti)$ system without Mo and W, the $L1_2 \rightarrow D0_{24}$ transformation requires a Ti concentration exceeding 8.0 at.% in the $\gamma'$ phase. The addition of 3 at.% W or Mo reduces this critical Ti concentration to 6.5 at.% and 6.6 at.%, respectively. When both W and Mo (in total 6 at.%) are added, the required Ti concentration further decreases to 4.5 at.%. These results demonstrate that Mo and W significantly promote the $L1_2 \rightarrow D0_{24}$ transformation in high-Ti alloys, supporting the conclusion that segregation of Ti, Mo, and W at SESFs drives local phase transformation toward the energetically favorable $D0_{24}$ structure. It is worth noting, however, that no grid-like ordering was observed along SESFs in alloy 8Ti by HAADF-STEM.

Similar behavior has been reported in low-Nb and high-Nb Ni-based superalloys (*Lilensten et al. 2021, Egan et al. 2022*). In low-Nb alloys, strong Nb segregation occurs along SESFs without producing grid-like ordering in HAADF-STEM images. By contrast, high-Nb alloys show both strong segregation and ordered grid-like $L1_2 \rightarrow D0_{24}$ transformation along SESFs. The η phase formed at SESFs suppresses the deleterious thickening of SESFs into microtwins by restricting Shockley-partial motion along SESF intereface. Additionally, η-forming elements, such as Ta and Nb, possess low diffusivity in Co- and Ni-based alloys (*Janotti et al. 2004, Neumeier et al. 2016*), further inhibiting microtwin formation. In this work, *ab initio* calculations indicate that segregation of Ti, Mo, and W at SESFs can also promote local phase transformation toward the η ($D0_{24}$) structure, but this transformation requires sufficiently high concentrations of Mo and W.

Although the refractory elements Mo and W in alloy 8Ti also possess very low diffusion coefficients (*Reed 2008, Janotti et al. 2004, Neumeier et al. 2016*), the accelerated diffusion of other solutes at 1223 K promotes SESF propagation, enabling SESFs to thicken into microtwins when no LPT occurs. Because η-type ordering does not form along SESFs in alloy 8Ti, SESFs can more readily undergo successive thickening into microtwins. This explains the abundant microtwins observed in alloy 8Ti.

Based on the analogy with low-Nb and high-Nb Ni-based superalloys discussed above, it is expected that increasing the W or Mo content in alloy 8Ti would promote grid-like ordering along SESFs, thereby stabilizing the $D0_{24}$ structure and suppressing SESF thickening into microtwins.

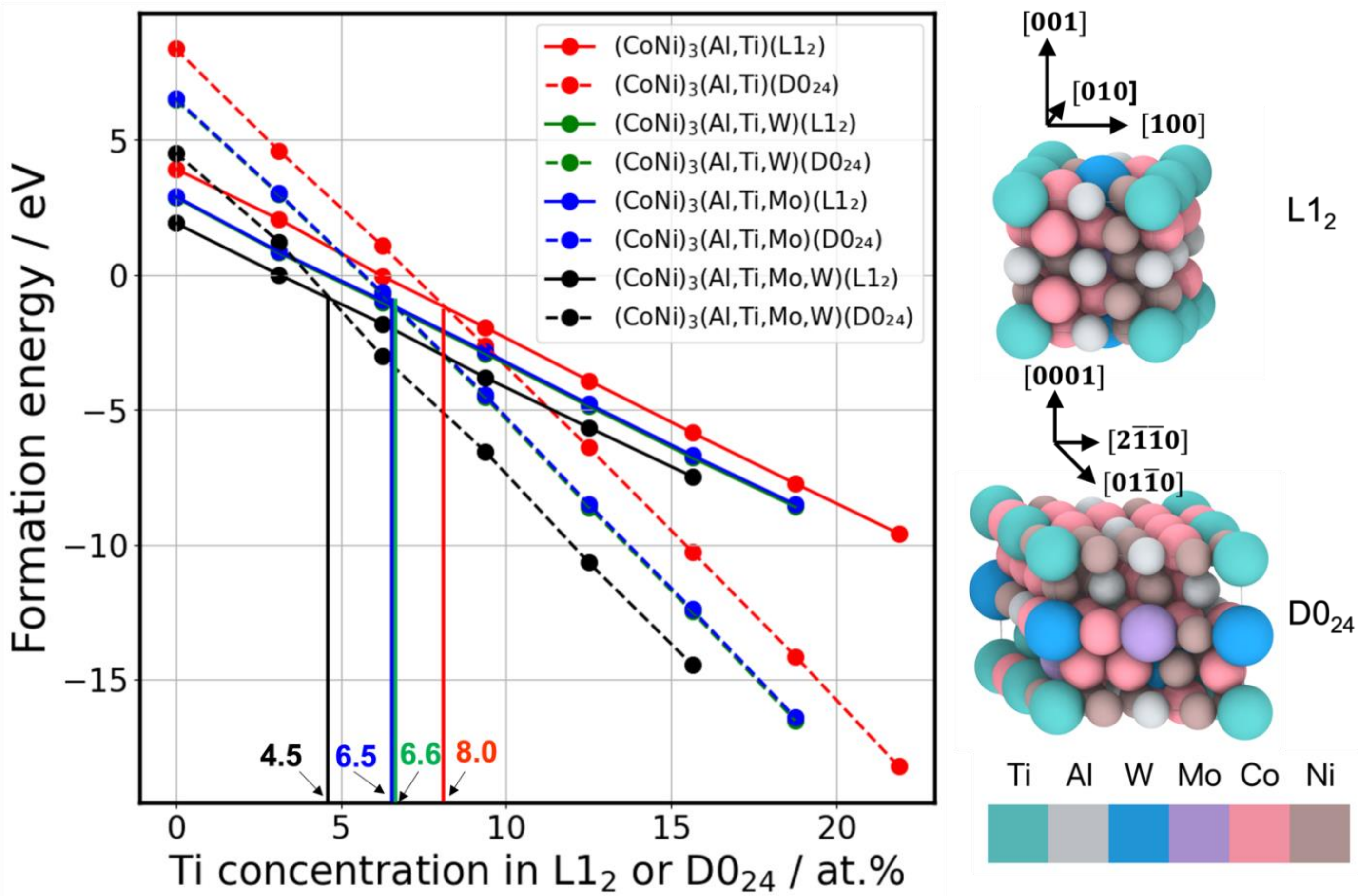

**Fig. 12.** Formation energies of the $L1_2$-type and $D0_{24}$-type $(Co_{37.5},Ni_{37.5})_3(Al_{25-x},Ti_x)$, $(Co_{37.5},Ni_{37.5})_3(Al_{22-x},Ti_x,Mo_3)$, $(Co_{37.5},Ni_{37.5})_3(Al_{22-x},Ti_x,W_3)$ and $(Co_{37.5},Ni_{37.5})_3(Al_{19-x},Ti_x,Mo_3,W_3)$ as a function of different Ti concentrations (Illustrations of the supercell models for $L1_2$ and $D0_{24}$ structures corresponding to the composition $(Co_{37.5},Ni_{37.5})_3(Al_{19-x},Ti_x,Mo_3,W_3)$ are shown.).

### 4.3 The effect of planar faults on creep resistance

#### 4.3.1 The shearing rate of planar faults in the γ′ precipitates

Numerous factors influence the creep resistance of the γ′ precipitates hardened superalloys, including the volume fraction of the γ′ precipitates, lattice misfit, planar faults energy and elements diffusivity. In this section, we focus primarily on the impact of planar faults on creep resistance. Under creep conditions, some dislocations readily shear the γ′ precipitates with low-energy APBs or SESFs, rather than bypassing the γ′ precipitates via dislocation climb. The moving velocity of the LPDs with planar faults ($v_{\text{planar}}$), dragged by Cottrell atmosphere, deciding average shear strain rate in the γ′ precipitates, can be estimated using the equation (*Hirth 1982*):

$$v_{\text{planar}} = \frac{\tau_{\text{app}} b_{\text{LPD}} D_{\text{solute}} k_{\text{B}} T}{C_0 \beta^2 \ln (R/r_0)} \tag{3}$$

where $\tau_{\text{app}} = \sigma_{\text{app}}/M$ is the resolved shear stress of LPD ($\sigma_{\text{app}}$ is the applied stress on samples at 1223 K and $M$ is the Taylor factor for the FCC polycrystalline matrix (*Courtney 2005*)), $b_{\text{LPD}}$ is the Burgers vector magnitude of the LPD, $D_{\text{solute}}$ is the diffusivity of the solute in Cottrell atmosphere, $k_{\text{B}}$ is the Boltzmann constant, $T$ is the temperature (K), $C_0$ is solute composition in Cottrell atmosphere and $\beta$ is a factor related to the interaction energy between dislocation and solutes (*Hirth 1982*), $R$ and $r_0$ are the outer cutoff radius and core radius of leading dislocation, respectively. The $R$ is usually equal to $10^4 \cdot r_0$ and therefore the $\ln(R/r_0)$ can be equal to a numerical factor 4.6 (*Hirth 1982, Cottrell and Jaswon 1949*). The element W is the slowest diffusion solute in three alloys. Therefore, the diffusion rate of W decides the LPD moving velocity. The values of all parameters ($\tau_{\text{app}}$, $\sigma_{\text{app}}$, $M$, $b_{\text{LPD}}$, $D_{\text{solute}}$, $k_{\text{B}}$, $C_0$, $\beta$ and $T$) are summarized in **Table 3**.

**Table 3.** Materials parameters used for dislocation mobility ($v_{\text{planar}}$) and dislocation density ($\rho_{\text{LPD}}$) calculation in the γ′ precipitates.

| **Parameters** | **Values** |
|---|---|
| Temperature, $T$ | 1223 K |
| Applied stress, $\sigma_{\text{app}}$ | 241 MPa |
| Boltzmann constant, $k_{\text{B}}$ | $k_{\text{B}} = 1.380649 * 10^{-23}\ \text{J} \cdot \text{K}^{-1}$ |
| Interdiffusion coefficient for W solutes at 1223 K, $D_{\text{solute}}$ | $D_{\text{solute}} = 1.2 \times 10^{-17}\ \text{m}^2 \cdot \text{s}^{-1}$ (*Cui et al. 2013*). |
| The resolved shear stress of LPD, $\tau_{\text{app}}$ | $\tau_{\text{app}} = \sigma_{\text{app}}/M = 78.8$ MPa |
| Taylor factor for the FCC polycrystalline matrix, $M$ | 3.06 |
| Burgers vector magnitude of the LPD, $b_{\text{LPD}}$ | $b_{\text{LPD}} = b_{a/2[110]} = 0.253$ nm in alloys 0Ti and 4Ti, $b_{\text{LPD}} = b_{a/6[112]} = 0.146$ nm in alloy 8Ti |
| Solute composition in Cottrell atmosphere, $C_0$ | $1.81 \times 10^{27}\ \text{m}^{-3}$ for 2.26 at.% W in LPD of alloy 0Ti, $2.05 \times 10^{27}\ \text{m}^{-3}$ for 2.56 at.% W in LPD of alloy 4Ti, and $1.15 \times 10^{27}\ \text{m}^{-3}$ for 1.44 at.% W in LPD of alloy 8Ti |

| Factor related to the interaction energy between dislocation and solutes, $\beta$ | $\beta$ = 1.9 × 10^-29^ N·m^2^ |
|---|---|
| Effective stress for microtwinning, $\tau_{\mathrm{eff}}^{\mathrm{twin}}$ | $\tau_{\mathrm{eff}}^{\mathrm{twin}} = \tau_{\mathrm{app}}$ = 78.8 MPa |
| Burgers vector magnitude of Shockley partials in microtwins, $b_{\mathrm{tp}}$ | $b_{\mathrm{tp}} = b_{a/6[112]}$= 0.146 nm |
| Average number of planar defects per precipitate, $n$ | $n$ = 9.0 in alloy 0Ti, $n$ = 3.0 in alloy 4Ti and $n$ = 14.1 in alloy 8Ti (Based on STEM observations) |
| The average dislocation line length of the shearing leading partial dislocations, $\bar{l}$ | $\bar{l} = d_{\gamma'}$ = 236 nm in alloy 0Ti, $\bar{l} = d_{\gamma'}$ = 282 nm in alloy 4Ti and $\bar{l} = d_{\gamma'}$ = 376 nm in alloy 8Ti |
| The single precipitate volume, $V$ | $V = \pi {d_{\gamma'}}^3/6$ |

The super partial dislocation $a/2[110]$ with APBs velocity ($\upsilon_{\mathrm{APBs}}$) in alloys 0Ti and 4Ti is estimated to be ~ 1.90 nm/s and ~ 1.67 nm/s, respectively. There are usually two leading Shockley partials in front of SESFs, therefore, the magnitude of Burgers vector for two leading Shockley partials with SESFs is $2\left|b_{a/6[112]}\right|$ = 0.292 nm. The shearing velocity of the two leading Shockley partials with SESFs ($\upsilon_{\mathrm{SESFs}}$) of alloy 8Ti is estimated to be ~ 2.66 nm/s which is approximately ~ 1.5 times higher than super partial dislocation ($\upsilon_{\mathrm{APBs}}$) in alloys 0Ti and 4Ti.

The twin propagation rate ($\upsilon_{\mathrm{twins}}$), which is related to the shear strain rate during creep microtwinning, can be estimated by (*Karthikeyan et al. 2006*):

$$\upsilon_{\mathrm{twins}} = \frac{D_{\mathrm{ord}} \cdot (b_{\mathrm{tp}}/x^2)}{ln[f_{\gamma'} \Delta E/(2 \cdot \tau_{\mathrm{eff}}^{\mathrm{twin}} \cdot b_{\mathrm{tp}} - f_{\gamma'} \cdot E_{\mathrm{tt}})]} \tag{4}$$

where $\Delta \mathrm{E} = E_{\mathrm{pt}} - E_{\mathrm{tt}} \cdot (D_{\mathrm{ord}}/x^2)$ is the measure of the reordering rate, $D_{\mathrm{ord}}$ is the interdiffusion coefficient (The interdiffusion coefficient of element W was employed.) for the reordering during microtwinning, $x = 2b_{a/2[110]}$ is the short range diffusion length, $E_{\mathrm{pt}}$ = 0.7 J·m^-2^ is the energy of the two layered pseudotwin, $E_{\mathrm{tt}}$ = 0.02 J·m^-2^ is the energy of a true twin, $b_{\mathrm{tp}}$is the Burgers vector magnitude of Shockley partials in microtwins, $f_{\gamma'} = 0.69$ is the volume fraction of the $\gamma'$ precipitates in alloy 8Ti, $\tau_{\text{eff}}^{\text{twin}} = \tau_{\text{app}}$ is effective stress for microtwinning in this work without secondary precipitates consideration. The twin propagation rate ($\upsilon_{\text{twins}}$) is approximately 2.76 nm/s, which is comparable to the SESF propagation rate. The shearing rate of planar defects, including SESFs and microtwins, in the 8Ti alloy are higher than the APBs observed in the 0Ti and 4Ti alloys. Consistent with these observations, Barba et al. (*Barba et al. 2018*) reported that the slowest precipitate shearing mechanism corresponds to APBs formation, followed by SISFs, whereas SESFs exhibit the highest fault lengthening rates in Ni-based superalloys.

#### 4.3.2 The planar faults contribution in creep strain

The total creep strain ($\varepsilon_{\text{total}}$) is the sum of creep strain in the $\gamma'$ precipitates ($\varepsilon_{\gamma'}$) and creep strain in the $\gamma$ matrix ($\varepsilon_{\text{matrix}}$). The planar fault shearing is the main contribution for creep strain in the $\gamma'$ precipitates. Here the total shearing partial dislocation density in the $\gamma'$ precipitates was determined by (*Hirth 1982, Titus et al. 2015*):

$$\rho_{\text{LPD}} = \frac{n\bar{l}}{V} \tag{5}$$

where $n$ is the average number of planar defects per precipitate, $\bar{l}$ is the average dislocation line length of the shearing partial dislocations, and $V$ is the single precipitate volume. The values of all parameters ($n$, $\bar{l}$ and $V$) are summarized in **Table 3**. The shearing dislocation density on average in alloys 0Ti, 4Ti and 8Ti is $\rho_{\text{LPD}} = 30.8 \times 10^{13}$ m$^{-2}$, $\rho_{\text{LPD}} = 7.2 \times 10^{13}$ m$^{-2}$ and $\rho_{\text{LPD}} = 19.1 \times 10^{13}$ m$^{-2}$, respectively. Based on shearing partial dislocation density in alloys 0Ti and 4Ti, the shearing events decrease with Ti concentration increasing, which is mainly related with higher APB energy in alloy 4Ti. The higher dislocation density in alloy 8Ti is attributed into lower energy SESFs shearing.

The calculated average shear strain, $\bar{\gamma}_{\text{planar}}$, from planar defects (including APBs and SESFs) accommodated in the $\gamma'$ precipitates, can be determined from the Orowan equation, such that (*Hirth 1982, Titus et al. 2015*):

$$\bar{\gamma}_{\text{planar}} = b_{\text{LPD}}\rho_{\text{LPD}}\bar{x} \tag{6}$$

where $\bar{x}$ is the average slip distance of the LPD ($\bar{x}$ is equal to precipitate diameter $d_{\gamma'}$). The shear strain from planar defects in the $\gamma'$ precipitates is $\bar{\gamma}_{\text{planar,0Ti}} = 1.84$ %, $\bar{\gamma}_{\text{planar,4Ti}} = 0.51$ % and $\bar{\gamma}_{\text{planar,8Ti}} = 1.72$ % in alloys 0Ti, 4Ti and 8Ti, respectively. Based on the Tresca criterion (*Titus et al. 2015*), the creep strain accommodated by planar defects in the $\gamma'$ precipitates equate to $\varepsilon_{\text{planar}} = \bar{\gamma}_{\text{planar}}/2$, which is 0.92 %, 0.26 % and 0.86 %, respectively. Therefore, the dislocation shearing in the $\gamma'$ precipitates with low APB energy contributes to higher creep strain than that with high APB energy. The dislocation shearing in the $\gamma'$ precipitates with SESFs formation contributes to higher strain than that with APBs formation. Then, the planar fault shearing contribution for creep strain in the alloy 4Ti is the lowest, followed by the alloy 0Ti and finally the alloy 8Ti. Base on this, the SESF formation with Shockley dislocation shearing is not beneficial for creep resistance. Moreover, the SESFs are 'embryo' for creep twins which usually makes big contribution for creep strain. Lenz et al. (*Lenz et al. 2019*) also summarized thickening of microtwins contributes to much bigger creep strain compared with planar faults, such as SESFs, SISFs and APBs.

The macroscopic creep strain ($\varepsilon_{\text{twin}}$) produced by microtwins can be defined as (*Egan et al. 2022, Unocic et al. 2011*):

$$\varepsilon_{\text{twin}} = \frac{N_{\text{twin}}\Delta_{ave}}{L}, \Delta_{\text{ave}} = b_{\text{tp}}\lambda_{\text{twin}} = b_{a/6[112]}\lambda_{\text{twin}} \quad (7)$$

where $\frac{N_{\text{twin}}}{L} = 1.3 \times 10^5\ m^{-1}$ is the density of twins along a test line, see **supplementary materials Fig. S11**, $\lambda_{\text{twin}}$ = 497 nm is the average thickness of twins, see **supplementary materials Table S3**. Displacement of a single twin ($\Delta_{\text{ave}}$) can be given by the product of thickness ($\lambda_{\text{twin}}$) in {111} planes and the Burgers vector magnitude of Shockley partials ($b_{\text{tp}}$). The thickness measurement of creep twins was taken from BF-STEM image (**Fig. 4(d)**). The roughly estimated creep strain contribution from creep twins is $\varepsilon_{\text{twin}}$ = 0.9 % in alloy 8Ti, as is shown in **supplementary materials Fig. S11** and **Table S3**. The creep strain ($\varepsilon_{\text{APBs}}$, $\varepsilon_{\text{SESFs}}$ and $\varepsilon_{\text{Twins}}$) accommodated by planar defects (APBs, SESFs, and Twins) and their contributions in total creep strain ($\varepsilon_{\text{total}}$) were shown in **Table 4**.

Based on this, the total creep strain contributed by planar defects (SESFs + Twins) is 1.76 % in alloy 8Ti, which contributes about 40.4 % in the total strain accumulated in the creep specimen. As discussed above, the superdislocation shearing contribution with high APB energy in the $\gamma'$ precipitates for total creep strain are lower than 10 %, such as alloy 4Ti, while the dislocation shearing contribution with SESFs and microtwins formation is higher than 40 % for creep strain, such as alloy 8Ti. Therefore, in order to avoid the larger shearing creep strain in $\gamma'$ precipitates at high temperature, the appropriate Ti concentration (we suggest 0 < Ti < 8 at.%) should be employed in future superalloys.

**Table. 4.** The creep strain ($\varepsilon_{\text{APBs}}$, $\varepsilon_{\text{SESFs}}$ and $\varepsilon_{\text{Twins}}$) accommodated by planar defects (APBs, SESF, and Twins), where $\varepsilon_{\text{planar}} = \varepsilon_{\text{SESFs}} + \varepsilon_{\text{Twins}}$ in alloy 8Ti, and their contributions in total creep strain ($\varepsilon_{\text{total}}$).

| Alloys | $\varepsilon_{\mathrm{APBs}}$, % | $\varepsilon_{\mathrm{SESFs}}$, % | $\varepsilon_{\mathrm{Twins}}$, % | $\varepsilon_{\mathrm{total}}$, % | $\varepsilon_{\mathrm{planar}}/\varepsilon_{\mathrm{total}}$, % |
|---|---|---|---|---|---|
| 0Ti | 0.92 % | - | - | 9.04 % | 10.2 % |
| 4Ti | 0.26 % | - | - | 4.24 % | 6.1 % |
| 8Ti | - | 0.86 % | 0.9 % | 4.36 % | 40.4 % |

### 4.4 Why does alloy 8Ti still exhibit the best creep resistance?

Based on the calculations in **Table 4**, the main contribution to the total creep strain in alloys 0Ti, 4Ti and 8Ti originates from dislocation mobility in the γ matrix, $\varepsilon_{\mathrm{matrix}}/\varepsilon_{\mathrm{total}} = 89.8$ % in alloy 0Ti and $\varepsilon_{\mathrm{matrix}}/\varepsilon_{\mathrm{total}} = 93.9$ % in alloy 4Ti and $\varepsilon_{\mathrm{matrix}}/\varepsilon_{\mathrm{total}} = 59.6$ % in alloy 8Ti, which are strongly affected by the volume fraction of the γ′ precipitates (*Yokokawa et al. 2020*) and the γ/γ′ lattice misfit (*Chang et al. 2018*). According to previous studies, the creep resistance of superalloys increases with increasing γ′ precipitate volume fraction and reaches an optimum when the γ′ volume fraction approaches ~ 0.70 (*Reed 2008, Yokokawa et al. 2020*). In alloy 8Ti, the γ′ volume fraction is $f_{\gamma'} = 0.69$, which is very close to this optimal value. In contrast, $f_{\gamma'}$ is 0.47 in alloy 0Ti and 0.53 in alloy 4Ti, respectively.

**Table 5.** Materials parameters used for dislocation mobility calculation in the γ matrix channel.

| **Parameters** | **Values** |
|---|---|
| Shear modulus of the γ matrix, $G_{\gamma}$ | 32 GPa at 1223 K |
| Young's modulus of the γ matrix, $E_{\gamma}$ | 83.2 GPa at 1223 K |
| Poisson's ratio, $\nu$ | 0.3 |
| Gas constant, $R$ | $8.3144598\ \mathrm{J \cdot mol^{-1} \cdot K^{-1}}$ |
| Jog density coefficient, $C_{\mathrm{j}}$ | 0.012 (*Zhu et al. 2012*) |
| Diffusivity pre-factor, $D_0$ | $10^{-4}\ \mathrm{m^2 \cdot s^{-1}}$ (*Zhu et al. 2012*) |
| Activation energy for lattice diffusion, $Q_{\mathrm{eff}}$ | $320\ \mathrm{kJ \cdot mol^{-1}}$ (*Zhu et al. 2012*) |
| Lattice diffusion coefficient, $D_{\mathrm{lattice}}$ | $D_{\mathrm{lattice}} = D_0 \exp\left(-\frac{Q_{\mathrm{eff}}}{RT}\right)$ |
| Volume fraction of the γ′ precipitates, $f_{\gamma'}$ | $f_{\gamma'}$ = 0.47 in alloy 0Ti, $f_{\gamma'}$ = 0.53 in alloy 4Ti and $f_{\gamma'}$ = 0.69 in alloy 8Ti |
| Minimum creep rate, $\dot{\varepsilon}$ | $\dot{\varepsilon}_{0\mathrm{Ti}} = 1.5 \times 10^{-5}\ \mathrm{s}^{-1}$ , $\dot{\varepsilon}_{4\mathrm{Ti}} = 1.2 \times 10^{-6}\ \mathrm{s}^{-1}$ and $\dot{\varepsilon}_{8\mathrm{Ti}} = 4.4 \times 10^{-7}\ \mathrm{s}^{-1}$ |
| Burgers vector magnitude of perfect dislocation in the γ matrix, $b_{a/2[110]}$ | 0.253 nm |
| The average dislocation network spacing (measured by STEM image), $d_{\mathrm{net}}$ | $d_{\mathrm{net}} = 51.1 \pm 12.7$ nm in alloy 4Ti<br>and $d_{\mathrm{net}} = 37.0 \pm 9.5$ nm in alloy 8Ti |
| The effective stress, $\sigma_{\mathrm{eff}}$ | $\sigma_{\mathrm{eff}} = 239$ MPa in alloy 0Ti, $\sigma_{\mathrm{eff}} = 144$ MPa in alloy 4Ti and $\sigma_{\mathrm{eff}} = 107$ MPa in alloy 8Ti |

The creep resistance of the γ matrix channels is strongly governed by the glide and climb velocities of dislocations. An explicit relationship between the dislocation mobility in γ matrix channels and microstructure (the volume fraction of the γ′ precipitate, the γ matrix channel width and the γ′ precipitate size) can be expressed by (*Dyson 2009, Zhu et al. 2012*):

$$v_m = \frac{2f_{\gamma'}(1-f_{\gamma'})(1/{f_{\gamma'}}^{1/3})C_{\mathrm{j}}D_{\mathrm{lattice}}\,sinh\left(\frac{\sigma_{\mathrm{eff}}{b_{a/2[110]}}^{2}L_{\mathrm{matrix}}}{\sqrt{6}k_{\mathrm{B}}TK_{\mathrm{CF}}}\right)}{b_{a/2[110]}} \tag{8}$$

where $b_{a/2[110]}$ is the Burgers vector magnitude ($a/2[110]$), $f_{\gamma'}$ is the volume fraction of the γ′ precipitates, $L_{\mathrm{matrix}}$ is the interspace between two neighboring γ′ precipitates, $C_{\mathrm{j}}$ is jog density coefficient, $D_{\mathrm{lattice}}$ is the lattice diffusion coefficient, $\sigma_{\mathrm{eff}}$ is the effective stress (The calculations provided in **Appendix 1**.), the constraint factor ($K_{\mathrm{CF}}$) accounts for the close proximity of the cuboidal particles, namely $K_{\mathrm{CF}} = 1 + 2f_{\gamma'}^{1/3}/3\sqrt{3\pi}(1 - f_{\gamma'}^{1/3})$ (*Zhu et al. 2012*). The values of all parameters ($\sigma_{\mathrm{eff}}$, $b_{a/2[110]}$, $L_{\mathrm{matrix}}$, $D_{\mathrm{lattice}}$, $C_{\mathrm{j}}$, $T$, $k_{\mathrm{B}}$, $f_{\gamma'}$ and $K_{\mathrm{CF}}$) are summarized in **Tables 3** and **5**. The interspace ($L$) between two neighboring γ′ precipitates is given by (*Dyson 2009*):

$$L_{\text{matrix}} = 1.6\, r_{\gamma'}\left(\left(\frac{\pi}{4f_{\gamma'}}\right)^{1/2} - 1\right) \tag{9}$$

where $r_{\gamma'}$ is the radius of the $\gamma'$ precipitate, which is $r_{\gamma'}$ = 118 nm in alloy 0Ti, $r_{\gamma'}$ = 141 nm in alloy 4Ti and $r_{\gamma'}$ = 188 nm in alloy 8Ti. From **Eq. (9)**, the average interspace ($L_{\text{matrix}}$) in alloys 0Ti, 4Ti and 8Ti is calculated to be 55.2 nm, 49.0 nm and 20.1 nm, respectively.

In superalloys, a significant part of creep life is spent at the secondary regime of creep with the minimum creep rate. Using **Eq. (8)** and the minimum creep rates, the dislocation mobile velocity at the secondary regime of creep is obtained as: $v_m =$ 790 nm/s for alloy 0Ti, $v_m =$ 7.9 nm/s for alloy 4Ti and $v_m =$ 1.3 nm/s for alloy 8Ti, respectively. At the secondary creep stage, the alloy 8Ti clearly exhibits the lowest dislocation mobility in the γ matrix. Notably, the dislocation mobility in alloy 0Ti is approximately two orders of magnitude higher than that in alloy 4Ti. This difference is primarily attributed to the absence of a forest dislocation network in alloy 0Ti, which otherwise would provide a network strength to impede dislocation motion, as evidenced by STEM observations (**Fig. 3**).

**Table 6.** Comparison of dislocation mobility in the γ matrix channels ($v_m$) and γ′ precipitates ($v_{\text{APBs}}$ and $v_{\text{SESFs}}$) of alloys 0Ti, 4Ti and 8Ti. (Microtwinning ($v_{\text{twins}}$) occurs in both the γ matrix and γ′ precipitates.)

| Alloys | $v_m$, nm/s | $v_{\text{APBs}}$, nm/s | $v_{\text{SESFs}}$, nm/s | $v_{\text{twins}}$, nm/s |
|---|---|---|---|---|
| 0Ti | 790 | 1.9 | - | - |
| 4Ti | 7.9 | 1.67 | - | - |
| 8Ti | 1.3 | - | 2.66 | 2.76 |

**Table 6** summarizes the dislocation mobility in the γ matrix channels and within the γ′ precipitates for alloys 0Ti, 4Ti, and 8Ti. In the APB-shearing dominated alloys 0Ti and 4Ti, the mobility of perfect dislocations in the γ matrix channels is significantly higher than that of superlattice partial dislocations in the γ′ precipitates. In contrast, in the SESF-shearing-dominated alloy 8Ti, the mobility of perfect dislocations in the γ matrix channels is lower than that of Shockley partial dislocations within the γ′ precipitates and microtwins. The narrow γ matrix channels and strong forest dislocation network in alloy 8Ti therefore provide strong resistance to creep deformation, whereas the low-energy SESF shearing mechanism weakens the creep resistance of the γ′ precipitates.

In addition, higher level of misfit stress (higher lattice misfit) is beneficial for creep resistance because they are associated with smaller dislocation network spacing which is supported by Neumeier et al. work (*Neumeier et al. 2021*) and Harada group work (*Murakumo et al. 2004, Zhang et al. 2005, Zhang et al. 2003*). The minimum creep rate has linear relationship with dislocation network spacings (*Zhang et al. 2003*). The lattice misfit increases with Ti concentration increasing at 1223 K, as HEXRD measurement results in **supplementary materials Fig. S12**. The lattice misfit is $\delta = 0.44$ % in alloy 0Ti, $\delta = 0.51$ % in alloy 4Ti and $\delta = 0.62$ % in alloy 8Ti, respectively. By HAADF-STEM investigation, however, in alloy 0Ti, there is no dislocation network appeared in the interface of the γ and $\gamma'$ phases. The main possibility is the most of dislocations choose to glide in the γ matrix channel and shear the $\gamma'$ precipitates. The dislocation network is one knitting process which requires more dislocation climb (*Huang et al. 2018, Parsa et al. 2025, Parsa et al. 2024*), which indicates less dislocations in alloy 0Ti have diffusion assisted climb. It influences the rafting process directly, leading to no rafting existing in alloy 0Ti. The rafted structure has been proved to increase the creep resistance of superalloys by more effectively hindering mobile dislocations from bypassing the $\gamma'$ precipitates (*Zhang et al. 2003*). This result leads to alloy 0Ti having much worse creep resistance than alloys 4Ti and 8Ti. The average dislocation network spacings in alloy 4Ti and 8Ti is measured as 51.1 ± 12.7 nm and 37.0 ± 9.5 nm, respectively. It is generally accepted that as the dislocation density increases within the γ channels, the accumulation of back stresses becomes significant. These back stresses originate from the elastic interactions among dislocations and act as an opposing force to further dislocation motion. With increasing creep strain, the back-stress field intensifies, making it progressively more difficult for additional dislocations to enter the channels. As a result, the effective dislocation mobility decreases, leading to a reduction in the creep rate. This can explain why the minimum creep rate is the lowest in alloy 8Ti ($\dot{\varepsilon}_{8\mathrm{Ti}} = 4.4 \times 10^{-7}\ \mathrm{s}^{-1}$), followed by alloys 4Ti ($\dot{\varepsilon}_{4\mathrm{Ti}} = 1.2 \times 10^{-6}\ \mathrm{s}^{-1}$) and 0Ti ($\dot{\varepsilon}_{0\mathrm{Ti}} = 1.5 \times 10^{-5}\ \mathrm{s}^{-1}$).

In high-Ti alloys such as 8Ti, the smaller γ matrix channel size combined with the higher $\gamma'$ precipitates volume fraction and higher lattice misfit results in: smaller dislocation network spacing and stronger resistance to dislocation mobility. Consequently, the pronounced strengthening effect from the high $\gamma'$ precipitates volume fraction and high lattice misfit in alloy 8Ti provides the lowest creep rate and ultimately accounts for its superior creep resistance compared with alloys 0Ti and 4Ti.

**4.5 Alloy design guidance**

Increasing the concentration of Ti, as γ' phase forming element, generally leads to a higher γ' volume fraction and a larger γ/γ' lattice misfit, both of which strongly influence creep resistance. A high γ' volume fraction reduces the dislocation mobility within the γ matrix channels, thereby effectively improving creep resistance. Additionally, a large lattice misfit generates significant misfit stresses that accommodate part of the applied stress, reducing the effective stress for creep. Increased lattice misfit also leads to a reduced dislocation spacing or a higher trapped dislocation density, resulting in enhanced network strength ($\hat{\sigma}_{\mathrm{net}}$), as described in **Appendix 1**. This effect lowers the effective stress ($\sigma_{\mathrm{eff}}$) acting on dislocations' mobility in the γ matrix channels during creep, according to **Eq. (8)**.

The γ' phase shearing induces the formation of very stable coherent boundaries, such as SESFs, APBs, SISFs and microtwins. However, element segregation locally changes the properties of coherent boundaries, such as the disordered APB and SESF regions. **Fig. 13** illustrates how segregation-assisted precipitate shearing occurs involving $a/2\,\langle 110\rangle$ perfect dislocations and $a/6\,\langle 112\rangle$ Shockley partial dislocations. A more disordered APB region forms, which will also help to expand the γ region by the wetting effect during the high-temperature creep process.

Segregation can decline the SESF energy significantly (*Eurich et al. 2015*) which allows more $a/2\,\langle 110\rangle$ dislocations to choose to shear γ' precipitates rather than to climb over it. Moreover, the SESFs are 'embryo' for creep twins which usually makes big contribution for creep strain. Lenz et al. (*Lenz et al. 2019*) also summarized thickening of microtwins contributes much bigger creep strain compared with other planar faults, such as SISFs and APB. Consequently, the compressive creep rate associated with microtwins and SESFs formation is typically higher than the tensile creep rate associated with SISF formation in the secondary and tertiary creep stages (c.f. **Fig. 1(a)** and **(b)** in ref. (*Lenz et al. 2019*)). In addition, the dislocation pile-up events often occur at the twin boundaries. These dislocations pile-ups can eventually cause stress concentrations which can lead to crack nucleation and propagation along the twin-parent interfaces (*Barba et al. 2017*). As a result, the appearance of microtwins is often associated with brittle fracture. Theoretical calculations further indicate that the migration velocity of planar defects such as SESFs and microtwins is higher than that of APBs. Compared with SESFs and microtwins, APBs therefore act as more effective barriers to precipitate shearing. To suppress γ' shearing via SESFs and microtwins and to enhance resistance to APB-mediated shearing at high temperatures, an appropriate Ti concentration should be carefully selected in CoNi-based superalloys.

Overall, owing to their high γ' volume fraction and large lattice misfit, high-Ti superalloys generally exhibit superior creep resistance. Nevertheless, to achieve an optimal balance among γ' volume fraction, lattice misfit, and the suppression of SESF formation and microtwinning, we propose that a relatively low Ti concentration should be employed in future CoNi-based superalloys. To compensate for the reduced γ' volume fraction, other γ' phase forming elements, such as Ta or Nb, may be added to enhance γ' precipitates and improve γ matrix channel creep resistance. Furthermore, the addition of Ta and Nb has been shown to increase the lattice misfit in CoNi-based superalloys (*Liang et al. 2023*), providing an additional pathway to improve high-temperature creep performance.

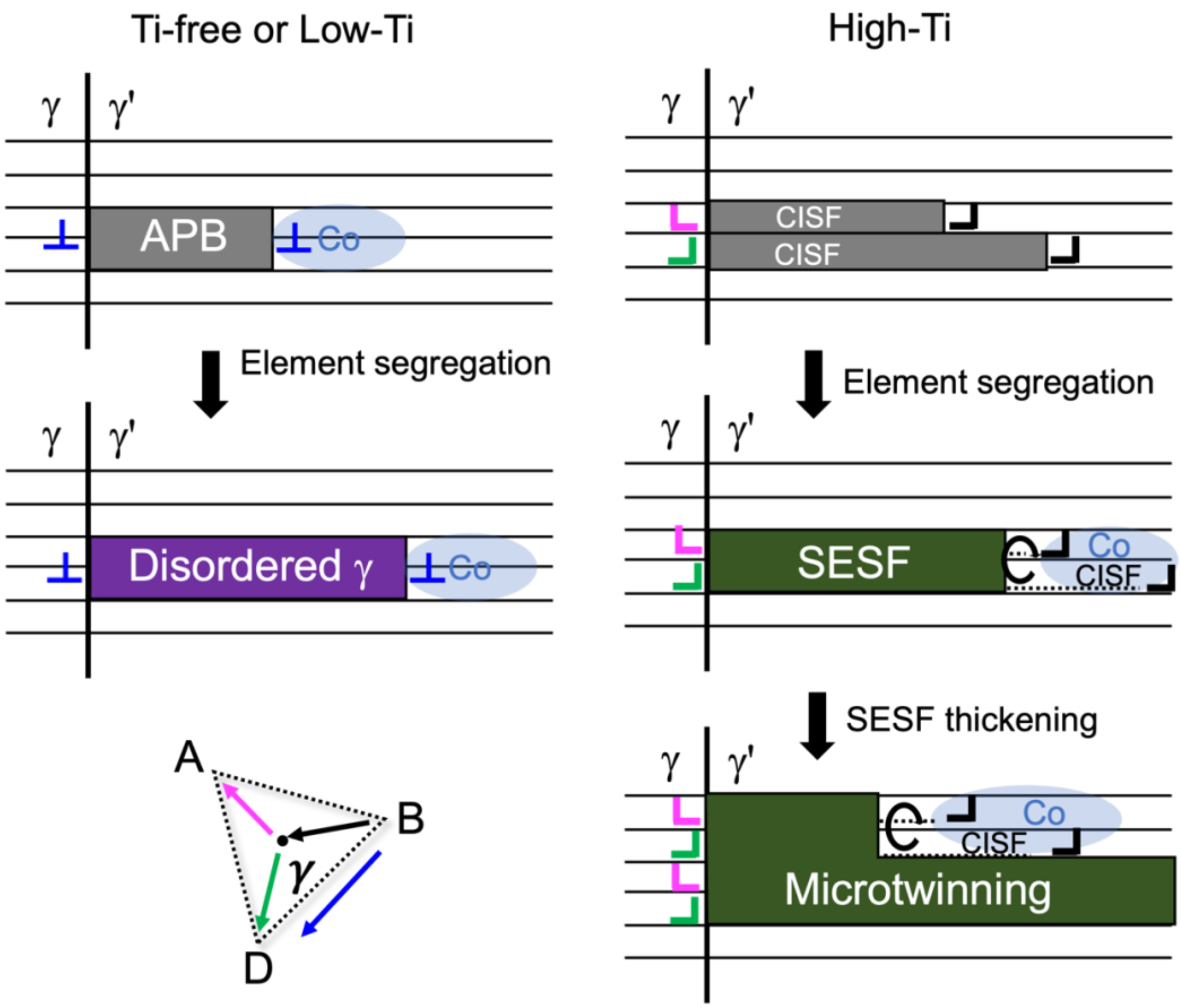

**Fig.13. Schematic summary of Ti-dependent planar-fault shearing modes:** disordered APBs in Ti-free and low-Ti alloys, and SESFs and microtwins in high-Ti alloys.

## 5. Conclusions

In this study, we studied shearing mechanisms in the γ' precipitates in CoNi based superalloys with different Ti contents during creep at 1223 K under low applied stress. The preferred shearing mode in γ' is governed by the relative magnitudes of the APB, CISF, and SESF energies, which determine the dominant planar-fault type. The main findings are as follows:

(1) The high Ti additions shift the dominant $\gamma'$ shearing mode from APBs in Ti-free and low-Ti superalloys to SESFs in high-Ti superalloys.

(2) In Ti-free or low-Ti superalloys, both the APB and CISF energies are lower compared with those in high-Ti alloys. Consequently, shearing of the γ' phase predominantly occurs via the motion of superlattice dislocations with Burgers vector $a/2\langle 110\rangle$, leaving behind a disordered APB region. Elemental segregation along these APBs on the (111) and (001) planes promotes local disorder transformation toward the γ phase.

(3) In high-Ti superalloys, the elevated APB energy suppresses this shearing mode with superpartial dislcoations, whereas the SESF energy decreases. The increased ratio $E_{\mathrm{CISF}}/E_{\mathrm{SESF}}$ facilitates the transformation of CISFs into SESFs. The high-energy CISFs, created by the passage of Shockley partials $a/6\langle 121\rangle$, are thus transformed into lower-energy SESFs through atomic reordering accompanied by elemental segregation at high temperature.

(4) Segregation of Ti, Mo, and W along SESFs in high-Ti superalloys can drive LPT toward ordered $D0_{24}$ structured η-type domains within the γ' precipitates based on our *ab initio* calculations. However, there is not grid-like ordering fromation along SESFs by HAADF-STEM observation which indicates there is no LPT phase transformation in alloy 8Ti. The absence of LPT along SESFs can not inhibit creep microtwinning. At high temperature, the diffusivity of refractory elements (Mo and W) becomes sufficiently high to promote SESF thickening. With increasing creep strain, thickened SESFs transform into microtwins.

(5) Microtwinning, including microtwins and SESFs, contributes most significantly to the overall creep strain, up to 40.4 % in alloy 8Ti, exceeding APBs in alloys 0Ti and 4Ti. Therefore, controlling the nucleation and thickening of microtwins is critical for improving creep resistance. A lower Ti content may be required to suppress microtwin formation under high-temperature creep conditions.

(6) Despite the higher microtwin fraction in the alloy 8Ti, its creep resistance is the best among the studied alloys. This superior performance can be attributed to its higher γ' volume fraction and larger γ/γ' lattice misfit, both of which strongly impede dislocation mobility and enhance creep resistance.

**Acknowledgements**

The author, Prof. Y.N. Cui, acknowledges the support from the National Natural Science Foundation of China under Grant Nos. 12222205 and 12172194, and from the Tsinghua University Initiative Scientific Research Program. The author, Dr. Zhida Liang, acknowledges the support from the National Natural Science Foundation of China under Grant Nos. 52501200. The authors gratefully acknowledge the scientific support and HPC resources provided by the Erlangen National High Performance Computing Center (NHR@FAU) of the Friedrich-Alexander-Universität Erlangen-Nürnberg (FAU) under the NHR project <Your NHR@FAU project ID>. NHR funding is provided by federal and Bavarian state authorities. NHR@FAU hardware is partially funded by the German Research Foundation (DFG) - 440719683.

**Appendix 1: The effective stress for dislocation glide in the γ matrix during creep.**

The effective stress during creep can be calculated by (*Dyson 2009*):

$$\sigma_{\rm eff} = \sigma_{\rm app} - \sigma_{\rm i} - \hat{\sigma}_{\rm net} \tag{A.1}$$

The internal stress ($\sigma_{\rm i}$) in the γ matrix has close relationship with creep rate ($\dot{\varepsilon}$) can be expressed by:

$$\dot{\sigma}_{\rm i} = \frac{f_{\gamma'}E_{\gamma}}{1-f_{\gamma'}}\left(1-\frac{\sigma_{\rm i}}{\bar{\sigma}_{\rm i}}\right)\dot{\varepsilon} \tag{A.2}$$

$$\bar{\sigma}_{\rm i} = \frac{2f_{\gamma'}}{1+2f_{\gamma'}}\sigma_{\rm app} \tag{A.3}$$

The 'network' strength ($\hat{\sigma}_{\rm net}$) results from dislocation network density ($\rho_{\rm T}$) with corresponding dislocation spacing ($\rho_{\rm T}^{-1/2}$):

$$\hat{\sigma}_{\rm net} = M\alpha G_{\gamma} b \rho_{\rm T}^{1/2} \tag{A.4}$$

$$\rho_{\rm T}^{1/2} = \frac{1}{d_{\rm net}} \tag{A.5}$$

where $G_{\gamma}$ is the shear modulus of the γ matrix (pure Co ($G_{0K} = 61.6$ GPa)), $E_{\gamma}$ is Young's modulus of the γ matrix, $d_{\rm net}$ is the dislocation spacing in dislocation networks, $\alpha$ is constant between 0 and 1 (we use 0.2 in this work), *b* is the Burgers vector magnitude for perfect dislocations in γ matrix (0.248 nm), The shear modulus of the γ matrix at 1223 K can be calculated based on $G = G_{\rm 0K} - 0.0009 * T - 0.000019 * T^2$ (GPa) (*Galindo-Nava et al. 2015*). The values for effective stress in three diffreent alloys (0Ti, 4Ti and 8Ti) are shown in **Table 5**. Since no dislocation networks are observed in the alloy 0Ti, we assume that the network strengthening contribution ($\hat{\sigma}_{\rm net}$) is equal to zero in this alloy.